\documentclass[manuscript]{aastex}

\shorttitle{GX 339-4: O/IR Monitoring for 2002-2010}
\shortauthors{Buxton et al.}

\begin{document}

\title{Optical and Near Infrared Monitoring of the Black-Hole X-ray Binary GX 339-4 During 2002-2010}

\author{Michelle M. Buxton, Charles D. Bailyn, Holly L. Capelo, and Ritaban Chatterjee}
\affil{Astronomy Department, Yale University, P.O. Box 208101, New Haven, CT 06520-8101}

\email{michelle.buxton@yale.edu}

\and

\author{Tolga Din\c{c}er and Emrah Kalemci}
\affil{Sabanci University, Orhanli-Tuzla, Istanbul, 34956, Turkey}

\and

\author{John A. Tomsick}
\affil{Space Sciences Laboratory, 7 Gauss Way, University of California, Berkeley, CA, 94720-7450, USA}

\begin{abstract}
We present the optical/infra-red lightcurve (O/IR) of the black hole X-ray binary GX 339-4 
collected at the SMARTS 1.3m telescope from 2002 to 2010.  During this time
the source has undergone numerous state transitions including hard-to-soft state transitions when we see large changes in the near-IR flux accompanied by modest changes in optical flux, and three rebrightening events in 2003, 2005 and 2007 after GX 339-4 transitioned from the soft state to the hard.  All but one outburst show similar behavior in the X-ray hardness-intensity diagram.  We show that the O/IR colors
follow two distinct tracks that reflect either the hard or soft X-ray state of the source.  Thus, either of these two X-ray states can be inferred from O/IR observations alone.  From these correlations we have constructed spectral energy distributions of the soft and hard states.  During the hard state, the near-IR data have the same spectral slope as simultaneous radio data when GX 339-4 was in a bright optical state, implying that the near-IR is dominated by a non-thermal source, most likely originating from jets.  Non-thermal emission dominates the near-IR bands during the hard state at all but the faintest optical states, and the fraction of non-thermal emission increases with increasing optical brightness.  The spectral slope of the optical bands indicate that a heated thermal source is present during both the soft and hard X-ray states, even when GX 339-4 is at its faintest optical state.  We have conducted a timing analysis of the light curve for the hard and soft states and find no evidence of a characteristic timescale within the range of 4-230 days.
\end{abstract}

\keywords{binaries: close --- black hole physics --- infrared: stars ---radiation mechanisms: non-thermal --- radiation mechanisms: thermal --- stars: individual (GX 339-4)}

\section{Introduction}

Observations of X-ray emitting binary stars provide clear evidence for
the presence of accreting black holes in some of these systems since 
dynamical measurements of the mass of the accreting compact object are 
greater than the maximum mass of a neutron star (Remillard \& McClintock 2006 and
references therein).  Around 20 black hole X-ray binaries (BHXBs) have been discovered in our Galaxy so far.  Observations of these objects are seen to have variable emission at all wavelengths due to changes in the mass accretion rate, the appearance/disappearance of a jet, and variations in the jet emission strength.  

GX 339-4 is unusual among BHXBs in that it has not been observed to go into
an optical/infrared ``quiescent''
state during which the flux from the companion star dominates over the flux from
the accretion flow.  Observations of the optical counterpart during its faintest optical and X-ray state \citep{sha01} do not show absorption lines in the optical spectrum, and identification of the companion star has remained elusive.  This has
complicated efforts to determine the mass and binary parameters of the system 
\citep[e.g.,][]{bux03}.  However, observations of emission features
associated with the irradiated companion star and accretion disk were successful in determining the orbital 
period and mass function of the source \citep{hyn03} so it is clear that
this source does contain a black hole.  

While the lack of a quiescent state has
complicated studies of the binary parameters of GX 339-4, its continuous accretion
has made it a key source for studying accretion phenomena.  To do so requires simultaneous multiwavelength data that covers the spectral energy distribution (SED) as broadly as possible.  The optical and near-infrared (O/IR) spectral region is particularly important in GX 339-4 \citep{cor02,rus06,hom05,cor09} as it appears to be the location of a cross-over between a non-thermal and thermal dominated spectrum.  This is also true for other BHXBs including XTE J1550-564 \citep{jai01,cor02b,rus10,cha11,rus11}, XTE J1118+480 \citep{hyn06,bro10} and 4U 1543-47 \citep{bux04}.

During the soft-state of GX 339-4, the spectral slope of the O/IR emission is positive suggesting that its  origin is thermal in nature.  The O/IR spectrum in BHXRBs during this state is expected to be dominated by a multicolor blackbody representing the accretion disk.  Such a spectrum consists of a Rayleigh-Jeans tail ($\alpha = 2$, where $F_\nu \propto \nu^\alpha$), a flattened profile ($\alpha \sim 1/3$ for a viscous heated disk), followed by the Wein tail.  To discern the temperature of the thermal source one needs to observe the turnover to the Wein tail.  This turnover has not been directly observed in GX 339-4 but simultaneous O/IR and UV observations of GX 339-4 \citep{cad11} show the flux increases from the infrared to the UV.  Studies of other BHXBs \citep{hyn02,hyn05,zur11} have revealed that the broadband SED increases from optical to UV wavelengths and, in some cases, seen to peak in the UV.  However, the UV flux determination is sensitive to the reddening estimate which can be frought with difficulties \citep{fit99} placing large errors on the peak location.  The optical spectral slope in many BHXBs during X-ray outbursts has been observed to be $0.5 \lesssim \alpha \lesssim 1.5$ \citep[][]{hyn05}.  However, the optical spectra were insufficient to determine whether the disk flux arose from irradiation from an X-ray source (central or vertically extended) or viscous heating in the outer disk.  For this one needs simultaneous UV data.  Not surprisingly, therefore, there is conflicting evidence as to whether the disk in GX 339-4 is irradiated or not during X-ray outbursts \cite[e.g.][]{hyn03,hom05,mar05,mai09,cad11}.  

In the hard state, the near-infrared spectral slope in GX 339-4 differs significantly from the optical and often \citep[though not always,][]{cor09} lines up with the radio spectrum \citep{cor02,hom05,rus06,cor09} indicating that the majority of near-infrared emission during the hard state may be non-thermal emission from a jet \citep{fen01,fen04,fen05}.  The location of the non-thermal power-law break is an important parameter as it is used to estimate the jet power \citep{fen06}. \cite{mot85} first showed that, during the hard state, the IR to X-ray SED was well explained by a single power-law fit with a slope of $\alpha = -0.58$.  Subsequently, \cite{cor02} constructed two SEDs with quasi-simultaneous radio, O/IR and X-ray data and suggested that the power-law breaks in the near-IR.  \cite{now05} found that, using 10 simultaneous radio/X-ray hard-state observations of GX 339-4, the hard-state SED could be fit with a doubly-broken power-law that breaks in the IR, and that the position of the break changes with X-ray flux.  Recent studies are now showing this to be the case.  Using simultaneous radio and SMARTS O/IR data taken in 2010, \cite{cor09} found that the radio-to-IR power-law break was located at frequencies higher than the $H$-band when GX 339-4 was in the hard state.  However, when GX 339-4 was transitioning from the soft to hard state, the break had to be located at frequencies lower than $H$-band as the O/IR data fluxes were well below that expected from the extrapolated radio spectrum.  \cite{gan11} obtained mid-IR (\emph{WISE}) with quasi-simultaneous UV (\emph{Swift}), optical (Faulkes Telescope South), near-IR (Rapid Eye Mount Telescope) and radio (ATCA) data, and found that during the hard state jet emission most likely dominated in the mid-IR and the power-law break was also located in this spectral region \citep[contrary to][]{cor09}. 

The hard-state optical spectrum typically remains thermal in its origin and is most-likely a viscously-heated accretion disk, an irradiated accretion disk and/or companion star, or a combination of these. 

Timing studies are valuable tools in determining the nature of emission during the soft and hard X-ray states.  Very little has been done during the soft state in GX 339-4.  \cite{mot85} found no sign of variability in the power-density spectrum (PSD) of soft state optical data.  QPO's ranging between 0.05 - 1 Hz have been observed in the PSD of hard state data \citep{mot85,gan09,gan10} that may be attributed to oscillations within the accretion disk.  Cross-correlation and auto-correlation functions obtained from hard state data show that optical emission lags X-ray emission by $\sim$ 150ms \citep{gan08} while the IR lags X-rays by $\sim$ 100ms \citep{cas10}.  These timescales are much too short to be explained by reprocessing of X-rays by the outer disk or companion star \citep[$\sim$ 25 seconds,][]{gan08} and suggest the dominant origin of hard state O/IR emission is much closer to the black hole, possibly the base of the jet and/or corona.  

O/IR data taken with the ANDICAM camera on SMARTS has proven to be a valuable resource for multiwavelength studies of GX 339-4 \citep[][Din\c{c}er et al., submitted]{hom05,tom08,cor09,cad11} and other XRBs \citep{bux04,kal05,rus06,mig07,oro09,cor10,raw11}.  

Using ANDICAM data of GX 339-4 \cite{hom05} found strong correlations between X-ray and O/IR fluxes that differed between the hard and soft states, indicating that the dominant flux source differed in each case.  The power-law slope of the near-IR/X-ray correlation during the hard state was very similar to that observed for the radio/X-ray correlation, suggesting that the majority of near-IR emission was from a jet, although other components most likely contributed to the overall O/IR flux as well.  During the soft state, \cite{hom05} showed a 15-20 day lag in variations between the X-ray and $H$-band fluxes, with the $H$-band leading the X-ray.  This time difference is consistent with a viscous time scale and strongly suggests that the O/IR emission during the soft state originates from an accretion disk.  However, since the ratio of X-ray to O/IR flux during the soft state was only a factor of 10 higher than during the hard state, it was thought the disk was either moderately heated or not heated at all.  


\cite{cor09} also used ANDICAM O/IR data of GX 339-4 to discover tight correlations between the X-ray and O/IR fluxes with branches that corresponded to the soft, intermediate, and hard X-ray states.  Comparison between the IR/X-ray and optical/X-ray correlations suggested that the near-IR fluxes are dominated by non-thermal emission while the optical is dominated by thermal emission in both hard and soft X-ray states.  

\cite{hom05}, \cite{cor09}, and other studies have used only part of our O/IR dataset of GX 339-4.  In this paper, we present the entire O/IR light curve of GX 339-4 from 2002-2010, inclusive, with the intent on making it public and available to the research community.  We have also analyzed the O/IR fluxes and SED to: 1) identify correlations between the near-IR and optical fluxes that are dependent upon X-ray state, such as that seen by \cite{hom05} and \cite{cor09}; 2) place limits on the SED power-law break in the near-IR during the hard state; 3) investigate the nature of the flux sources during both the soft and hard X-ray states; and 4) perform a timing analysis of the O/IR data to identify any characteristic timescales that may be present.  In Section \ref{obs}, we describe our data reduction and photometric calibration methods, and present the O/IR light curve of GX 339-4 and an X-ray hardness-intensity diagram (HID).  In Section \ref{oirstates}, we distinguish two different patterns of O/IR behavior that are associated with the soft and hard X-ray states.  In Section \ref{seds}, we study the O/IR SEDs and compare them to phenomenological models representing thermal and non-thermal emission.  In Section \ref{timing}, we investigate the variabilility timescales in the O/IR data during both soft and hard X-ray states.  We discuss our results in the context of other studies of GX 339-4, and give conclusions, in Section \ref{discuss}.  

\section{Observations and Data Analysis} \label{obs}

\subsection{Optical and Near-Infrared}

O/IR observations of GX 339-4 were taken with the 
ANDICAM\footnote{http://www.ctio.noao.edu/noao/content/andicam} camera \citep{dep03} on the SMARTS 1.3m telescope \citep{sub10}. 
ANDICAM is a dual-channel imager that contains a movable internal mirror in the IR channel that allows
dithered images to be taken while a single optical exposure is undertaken.
We obtained pairs of images in $V+J$ and $I+H$ filters on a daily 
or near-daily basis whenever GX 339-4 was available in the night sky.  
The optical images were single exposures of 300 seconds.
The IR images contained 8 dithered images of 30 seconds each. The optical data were bias- and overscan-subtracted, and flat-fielded using the 
CCDPROC task in IRAF.  Infrared data were reduced using in-house IRAF scripts that 
flat-fielded, sky-subtracted, and combined dithered images.

\subsubsection{Magnitude Determination and Calibration}\label{magcal}

Uncalibrated O/IR magnitudes were determined using DAOPHOT in IRAF, for which the 
full-width-half-maximum, sky background and point-spread function parameters were 
determined separately for every image.  

Optical primary standard stars \citep{lan92} were used to calibrate the optical magnitudes 
of three comparison stars in the field of GX 339-4. For 210 photometric nights we calculated the 
calibrated magnitude of each comparison star using photometric zeropoints that were measured from the
SMARTS observations of the primary standard stars for each filter, 
correcting for atmospheric extinction derived from the standard star observations.  
The average magnitude of these three comparison 
stars was then used as a basis for differential photometry with respect to GX 339-4 
for all observations.  The near-infrared 
photometry was calibrated using near-infrared primary standard data \citep{per98} and two 
comparison stars were used as secondary standards.  Only 21 nights of photometric data were used to 
measure the calibrated magnitudes of the secondary standards as we had limited data on the near-infrared photometric zeropoints.  

\begin{figure}
\plotone{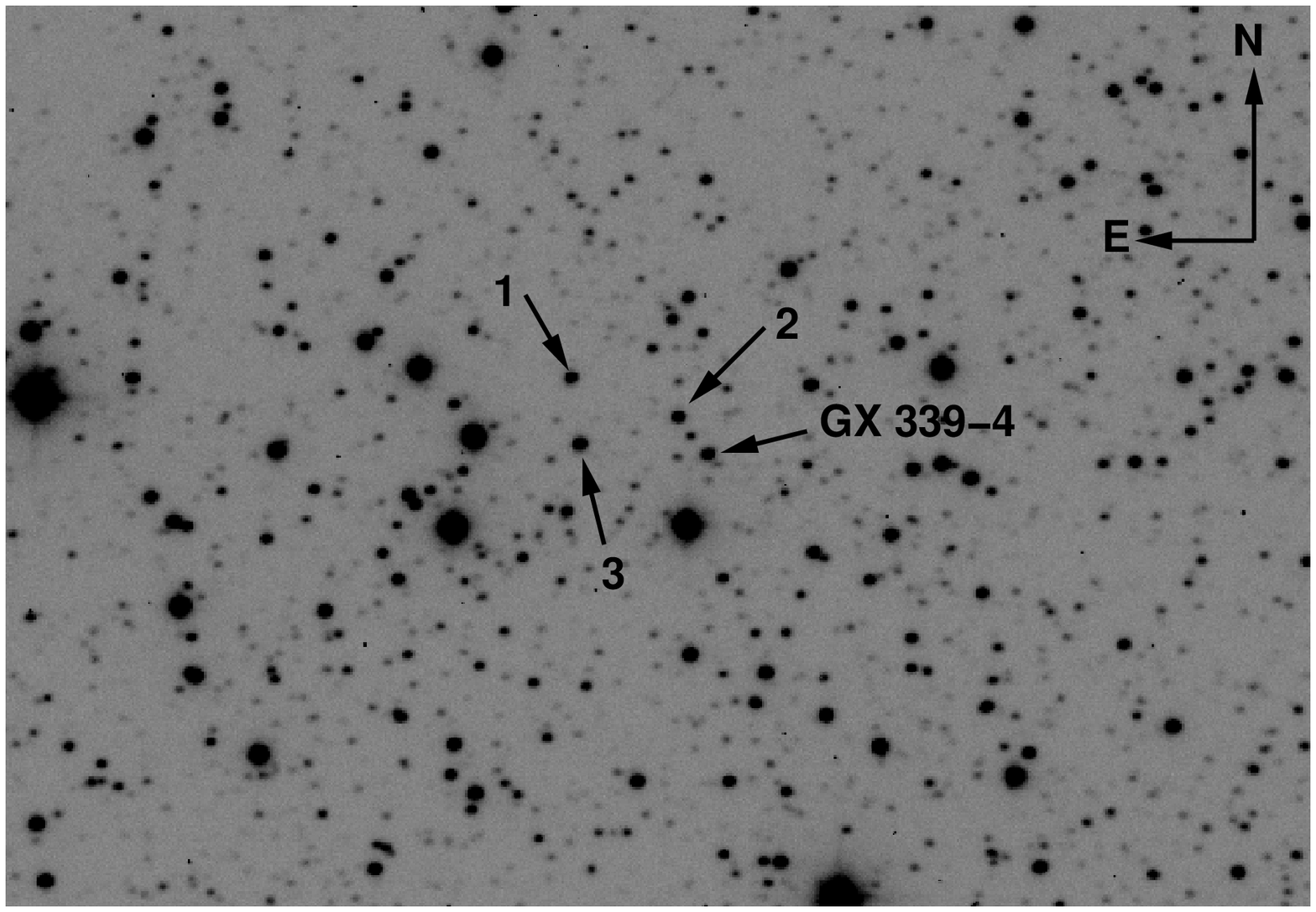}
\caption{V-band finding chart for GX 339-4.  Field of view is 3.7' x 2.7'.  }
\label{opfc}
\end{figure}

\begin{figure}
\plotone{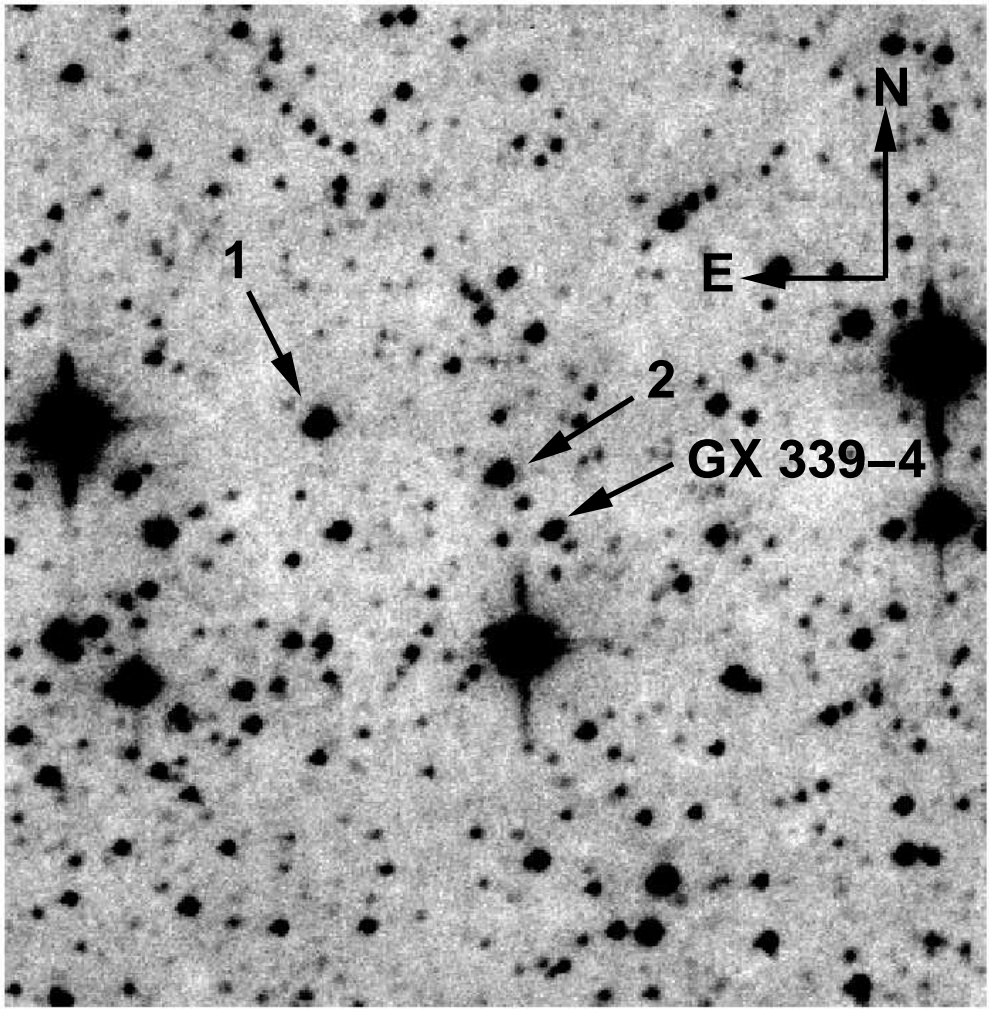}
\caption{H-band finding chart for GX 339-4, field size is 2.4' x 2.4'}
\label{irfc}
\end{figure}

We show the comparison stars in the
optical and near-infrared finding charts shown in Figures \ref{opfc} and \ref{irfc}.  
Our calibrated magnitudes of these comparison stars are given in Tables \ref{compstarsop} and \ref{compstarsir}.  
We also present the O/IR magnitudes of GX 339-4 in Table \ref{alldata}, the full version of which can be accessed online.

\subsubsection{Color correction}\label{colorcorr}

The ($V-I$) color of our secondary standards are given in Table \ref{compstarsop}, and they range from 1.37 - 2.28. For GX 339-4, the majority of ($V-I$) data vary over the X-ray states between 0.9 and 2.0.  The secondary standard colors overlap most of the ($V-I$) color range exhibited by GX 339-4.  We calculated the mean color-term coefficient for the $I$-band for 1052 photometric nights between August 2003 and December 2010 to be 0.062 $\pm$ 0.048 (1-$\sigma$).  Therefore, using the average color-term correction for the $I$-band, the color-term correction for the secondary stars ranges between 0.085 and 0.141 mag.  

Color correction in the near-infrared is minimal and, as such, is not required.

\subsubsection{Photometric Errors}\label{photerrs}

Errors on the $V$- and $I$-band differential photometry 
were obtained by calculating the 1-$\sigma$ error of 93 stars from 40 nights of data obtained
during 2006. 
Results that deviated more than $\pm$ 3-$\sigma$ from the mean were rejected, 
and the mean and $\sigma$ were recalculated until no more rejections were made.  The 1-$\sigma$ error for 
each star was plotted against their calibrated magnitude and a 4th-order polynomial was fitted to 
the resulting distribution.  Again we rejected any data that deviated more than
$\pm$ 3-$\sigma$ from the polynomial and recalculated the polynomial fit until no more rejections were made.
The same procedure was performed for the IR data, using 44 stars for the $J$-band and 29 stars for
the $H$-band, from 40 nights in 2006.  We used fewer
stars than the optical bands because the field of view is smaller 
and there are fewer stars to choose from.  

\begin{figure}
\plotone{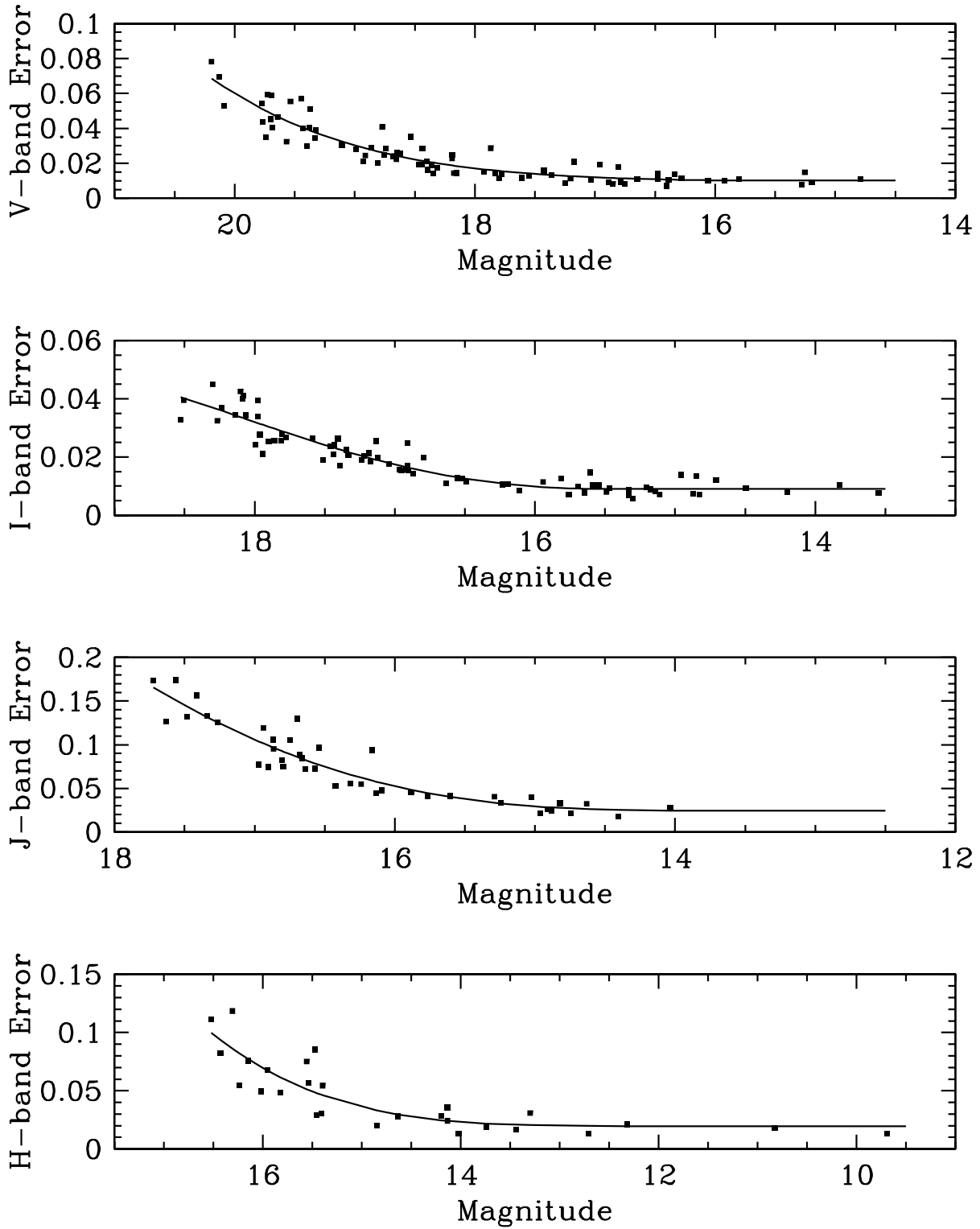}
\caption{Photometric error of stars in the GX 339-4 field versus magnitude for optical and near-infrared bands. The solid line is a 4th-order polynomial fit. The $y$-axis is the 1-$\sigma$ dispersion of measurements of stars as described in Section \ref{photerrs}.}
\label{photerr}
\end{figure}

Figure \ref{photerr} shows 1-$\sigma$ errors versus magnitude for the 
O/IR and the corresponding polynomial fits.  Once the polynomial reached a minimum, we
extended that constant numerical value to brighter magnitudes.  Thus
the optical photometric errors are 0.01 mag for $V \le 17.50$ and $I \le 16.50$. 
Infrared photometric errors are 0.02 mag for $J \le 14.10$ and $H \le 14.10$.  We refer the reader to Figure \ref{photerr} to ascertain errors on fainter magnitudes.

We also measured the error in calibrating the secondary star magnitudes by calculating the standard 
error of the mean 
over the number of photometric nights mentioned above.  The photometric calibration errors are as follows:  
$V_{err}$ = 0.003 mag, $I_{err}$ = 0.004 mag, $J_{err}$ = 0.02 mag, $H_{err}$ = 0.02 mag.
These values do not account for systematic errors associated with such effects as the difference in
effective filter responses between SMARTS and the standard system.  Such systematics are likely to 
contribute a few hundredths of a magnitude of calibration error.  Therefore we take the optical 
calibration errors
to be $\approx 0.02$ magnitudes and the IR calibration errors to be $\approx 0.03$ magnitudes. 

We note that in the majority of images the two nearby stars to GX 339-4 \citep[denoted as `A' and `B' in][]{sha01} contribute negligibly to the overall flux of GX 339-4 during its faintest optical state as DAOPHOT is able to extract the PSF profiles of both GX 339-4 and stars `A' and `B' (the latter two are combined into one PSF profile).  However, there are cases where the seeing or the image quality are degraded such that DAOPHOT was not able to separate the PSF profile of GX 339-4 from the two nearby stars.  For completion, we have included this data in Table \ref{alldata} but have denoted it with an asterix.

\subsection{X-ray Data}

The X-ray data from proportional counter array (PCA) onboard the Rossi
X-ray Timing Explorer (RXTE) were reduced using HEASOFT V6.7. After the
standard data screening, the background corrected count rates were extracted
from proportional counter unit 2 in 3 energy bands corresponding to 2.8-5.2 keV,
5.2-9.8 keV, 9.8-25 keV.

\subsection{O/IR Light Curve}

\begin{figure}
\plotone{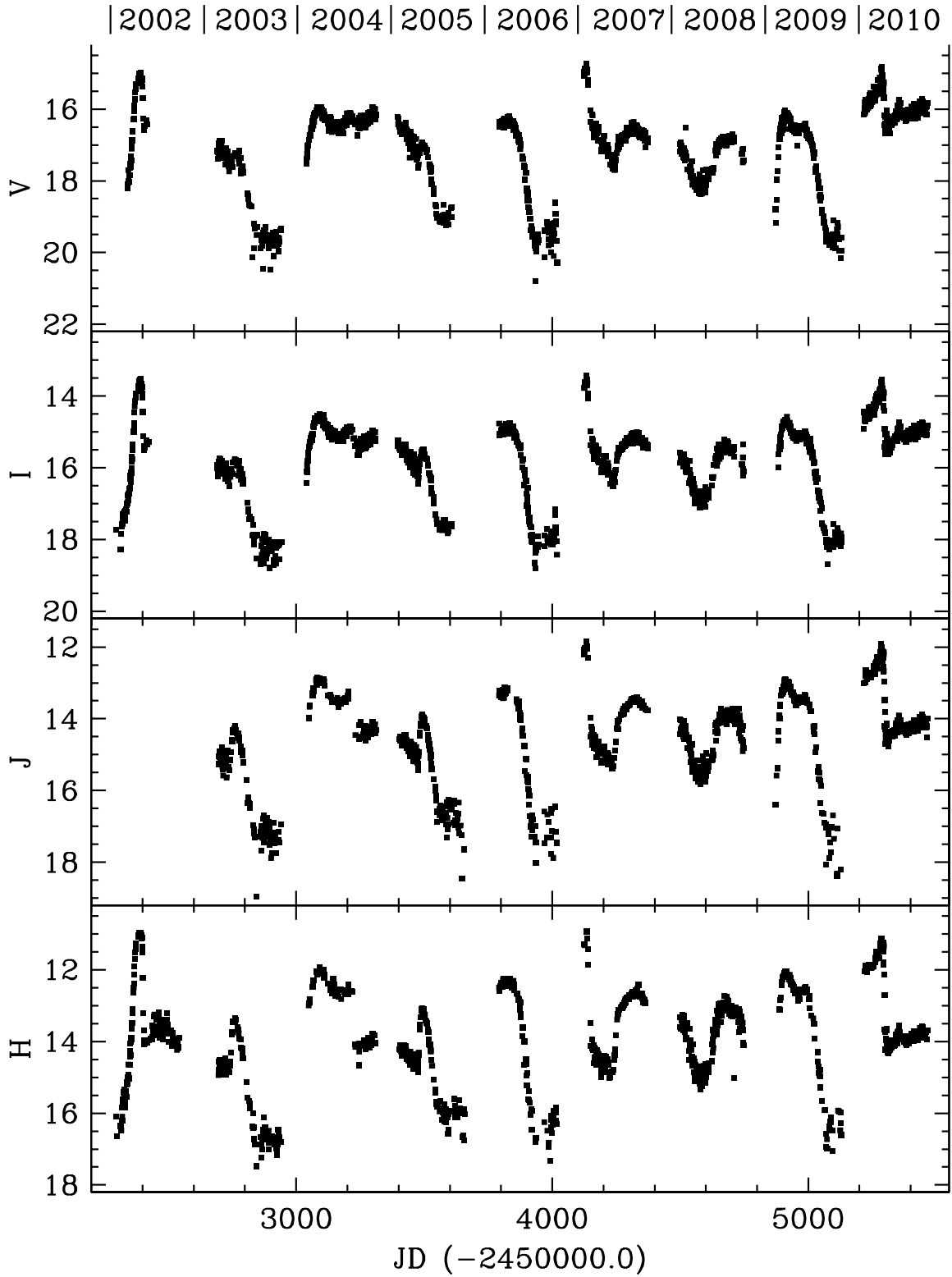}
\caption{The SMARTS 1.3m optical and near-infrared light curve of GX 339-4 during 2002-2010, inclusive.}
\label{oirlc}
\end{figure}

In Figure \ref{oirlc} we present the O/IR lightcurve of GX 339-4 during 2002-2010, inclusive. Gaps in the data occur when the source is behind the Sun. 

There are a number of interesting
features of this light curve. GX 339-4 clearly does not follow the pattern of long
periods of quiescence punctuated by outbursts with a 
fast-rise, exponential-decay morphology as is often seen in dwarf novae \citep[e.g.][]{can98} and
neutron-star X-ray binaries, such as Aql X-1 \citep{mai08} and many black hole
binaries \citep{che97}.  
Rather, the source is continually active with strong variability, featuring rapid rises
with a large range of decay times.  

In many cases, the
changes seen in the optical bands have smaller amplitudes than those that
occur at the same time in the near-infrared bands.  Obvious examples occur in
2003 and 2005 during the rebrightening phases, discussed in Section \ref{rebright}.  During these events an increase in flux is barely noticeable
in the $V$- or $I$-bands, but is more obvious in the $J$- and $H$-bands. 
There are also examples of sudden drops in IR flux (e.g. mid 2002, mid 2004 and early 2010) which are accompanied by more modest
changes in the optical \citep{bai04}.  These drops occur when the source transitions from the hard state into the soft state.  
This behavior is contrary to
what is expected if the O/IR flux originates solely from a thermal source, such as an accretion
disk.  Disk instabilities
\citep{can95,las01} and irradiation models \citep{dub99,dub01} generally attribute variability to
changes in temperature of the accretion disk.  In such cases, high luminosity should
be correlated with blue color.  However, this basic relationship clearly does not
hold for GX 339-4, which suggests that a non-thermal emission component is often
dominant.  It is, therefore, desirable to study the O/IR SED to ascertain what flux sources are present and how they vary and contribute to the overall flux during these changes in the light curve.  We present our O/IR SEDs in Sections \ref{softsed} and \ref{hardsed}.

\subsubsection{Rebrightening Events}
\label{rebright}

\begin{figure}
\plottwo{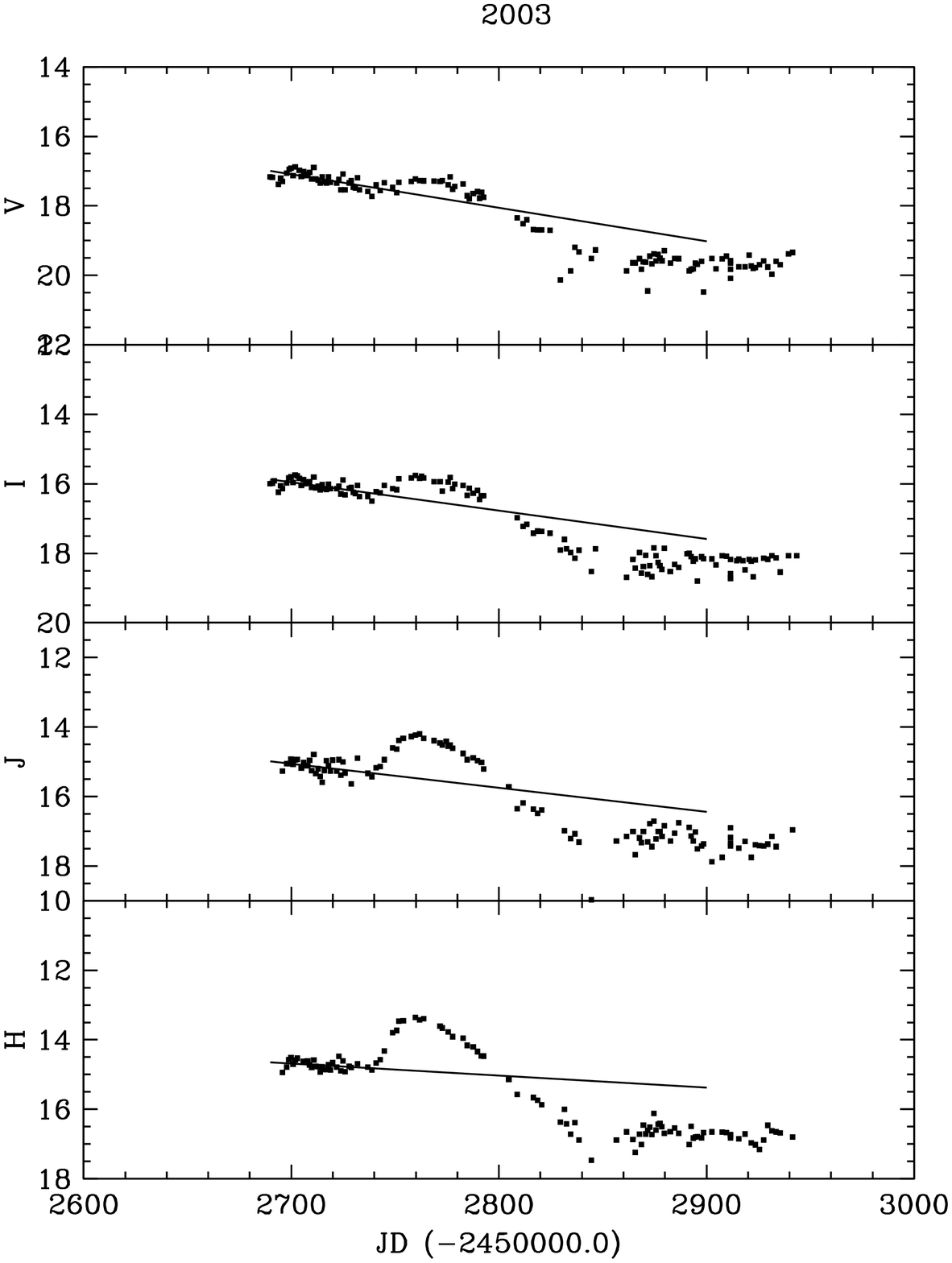}{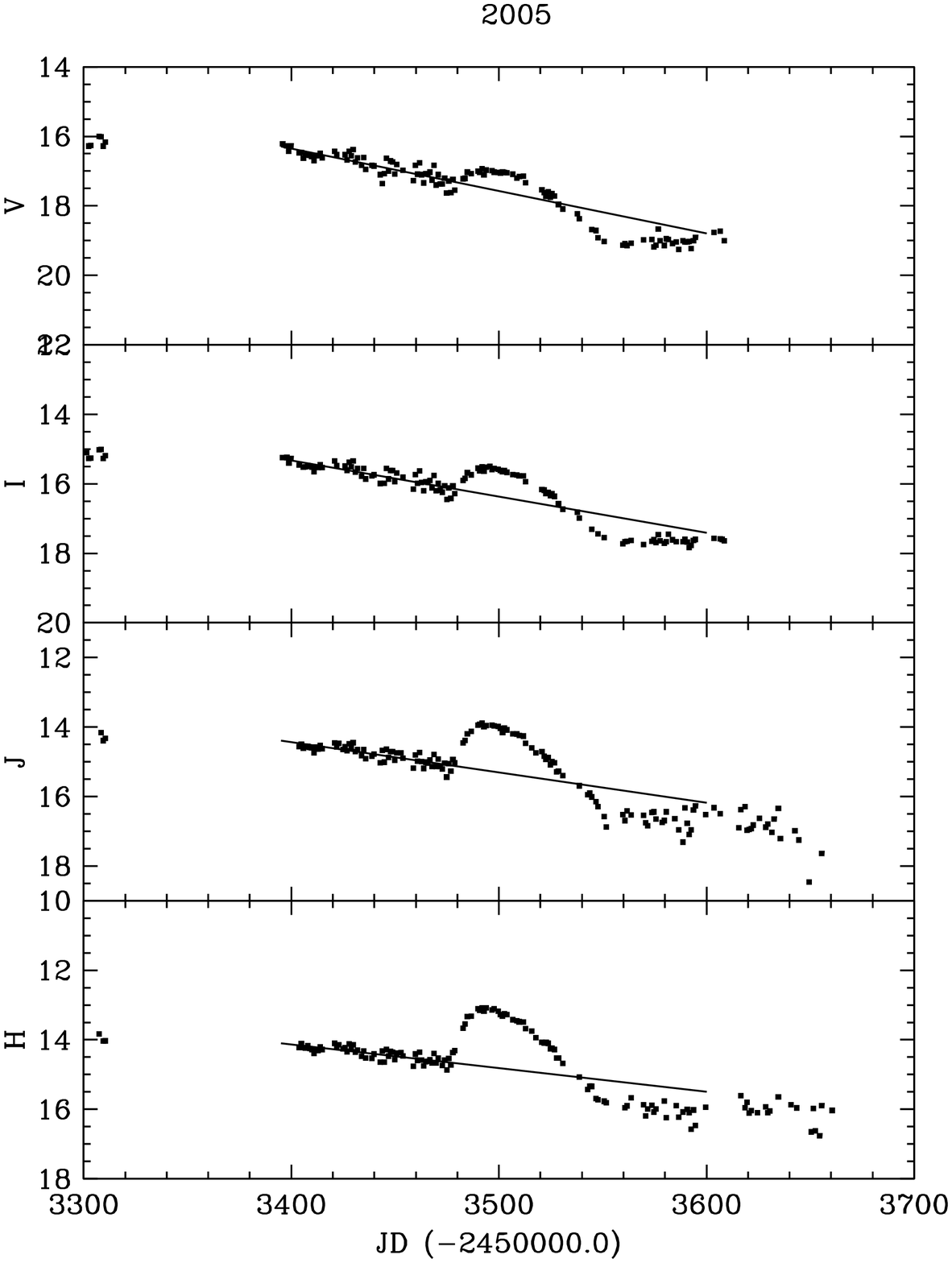}
\caption{O/IR light curves of GX 339-4 during 2003 and 2005, highlighting the rebrightening events.  The solid line is a best-fit line to the data immediately before the rise of the rebrightening event.  The slopes of these lines are given in Table \ref{rebrights_slopes}.}
\label{rebrights}
\end{figure}

During 2003 and 2005,
GX 339-4 displayed a slowly fading decay interrupted by a rebrightening which subsequently
declined back, but not to the extrapolation of the previous decay.  In Figure \ref{rebrights} 
we show these rebrightening events along with our best-fit straight lines to the data immediately 
before the rebrightening.  The slopes of our fits are given in Table \ref{rebrights_slopes}.  
Within errors, the O/IR decayed at the same rate between 2003 and 2005 with the exception of the $H$ band
for which the decay rate had slightly increased.  We extended the best-fit line in Figure \ref{rebrights} 
to highlight that the decay rate after the rebrightening event is not a continuation of that before, but
rather the O/IR flux continues to fade to a fainter, steady magnitude. 

The duration of the rebrightening event in 2003 ($\sim$ 100 days) is longer than in 2005 ($\sim$ 80 days), and the magnitude reached after the rebrightening event ended is slightly brighter in 2005 than in 2003 by $\sim$ 0.5 mags. This could mean that more material was accreted by the black hole and/or ejected via collimated jets in 2003 than in 2005.  

In 2007, GX 339-4 once again transitioned from the soft state to the hard state and experienced a rebrightening in O/IR fluxes.  However, the behavior after the rebrightening was markedly different to that in 2003 and 2005.  GX 339-4 did not fade back to a quiescent magnitude but remained in the hard state during 2008 until the end of 2009, fluctuating significantly in O/IR flux during that time. It is unclear why this should be the case.

With observations of future outbursts in GX 339-4 we will explore further the similarities and differences between these rebrightening events and how they relate to the overall outburst properties.  In addition, the decomposition of disk/jet flux during these events will be pursued in a future paper (also see Din\c{c}er et al., submitted).

\subsubsection{Hardness Intensity Diagram}
\label{hidsec}

\begin{figure}
\plotone{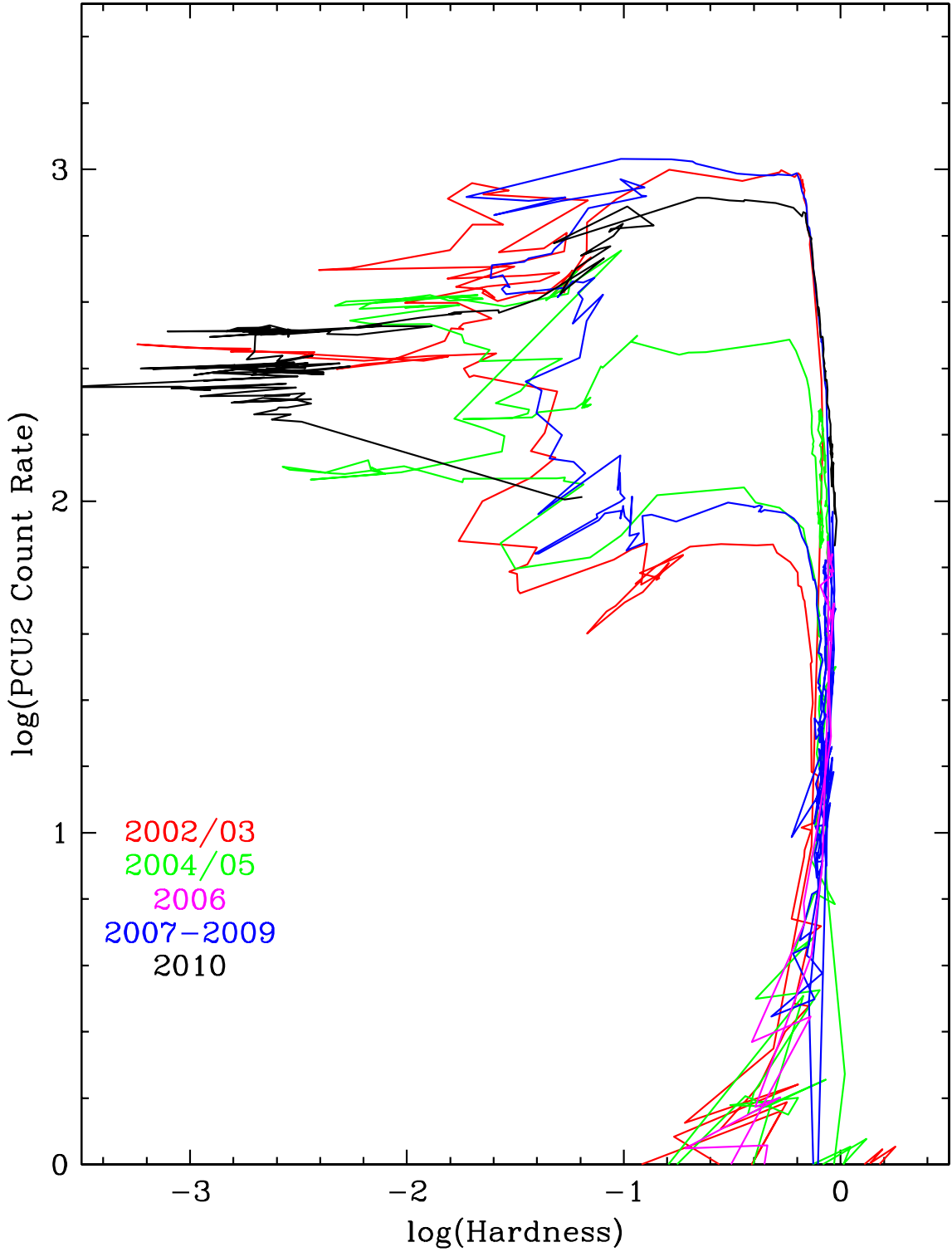}
\caption{Hardness-intensity diagram of GX 339-4 for 2002-2010.  Each separate outburst is shown in different colors.}
\label{hid}
\end{figure}

A useful tool in discerning when BHXRBs are in various states is the hardness-intensity diagram (HID) \citep{miy95} where X-ray hardness is plotted against the count rate.  Nearly all black-hole transients trace a `q-shaped' path counterclockwise in this diagram \citep{bel10}.  One exception is 4U 1630-47 \citep{tom05} during it's 2002-2004 outburst (although it did exhibit the q-shaped pattern in 1998).  In Figure \ref{hid} we show the HID of GX 339-4 for 2002 through 2010.  HIDs of the 2002/03, 2004/05, and 2007 outbursts have already been published in \cite{bel05}, \cite{hombel05}, and \cite{bel10}.  The 2008/09 data are an extension of the outburst started in 2007. The 2006 and 2007-09 outbursts are interesting for the following reasons.  In 2006, GX 339-4 experienced an outburst relatively short compared to the others and that never left the hard state. In the HID we see this outburst only resides in the lower vertical branch of the `q' and never resides in the loop.  However, we see in Figure \ref{oirlc} that the O/IR flux did increase significantly but did not peak at the same fluxes as in 2002, 2004, 2007 and 2010.  As already noted, the 2007-2009 outburst remained in the hard state for a very long time after transitioning from the soft state, and in the O/IR we see it fluctuating signficantly during this time.  The HID for this outburst does not show any obvious differences to those of other outbursts.  The only difference being the amount of time it has spent on the lower horizontal branch, also known as the hard-intermediate state \cite{bel10}.  The 2010 outburst looks to be typical of other outbursts in the HID.

\section{O/IR States in GX 339-4} \label{oirstates}

\subsection{O/IR Correlations}
\label{oircorrs}

\citet{cor09} described correlations between the X-ray and $V$- and $H$-band fluxes in GX 339-4 in which a clear delineation of data is seen between the hard, soft, and intermediate X-ray states. Here we investigate correlations within the O/IR and show how this relates to the X-ray behavior.  

\begin{figure}
\epsscale{.90}
\plotone{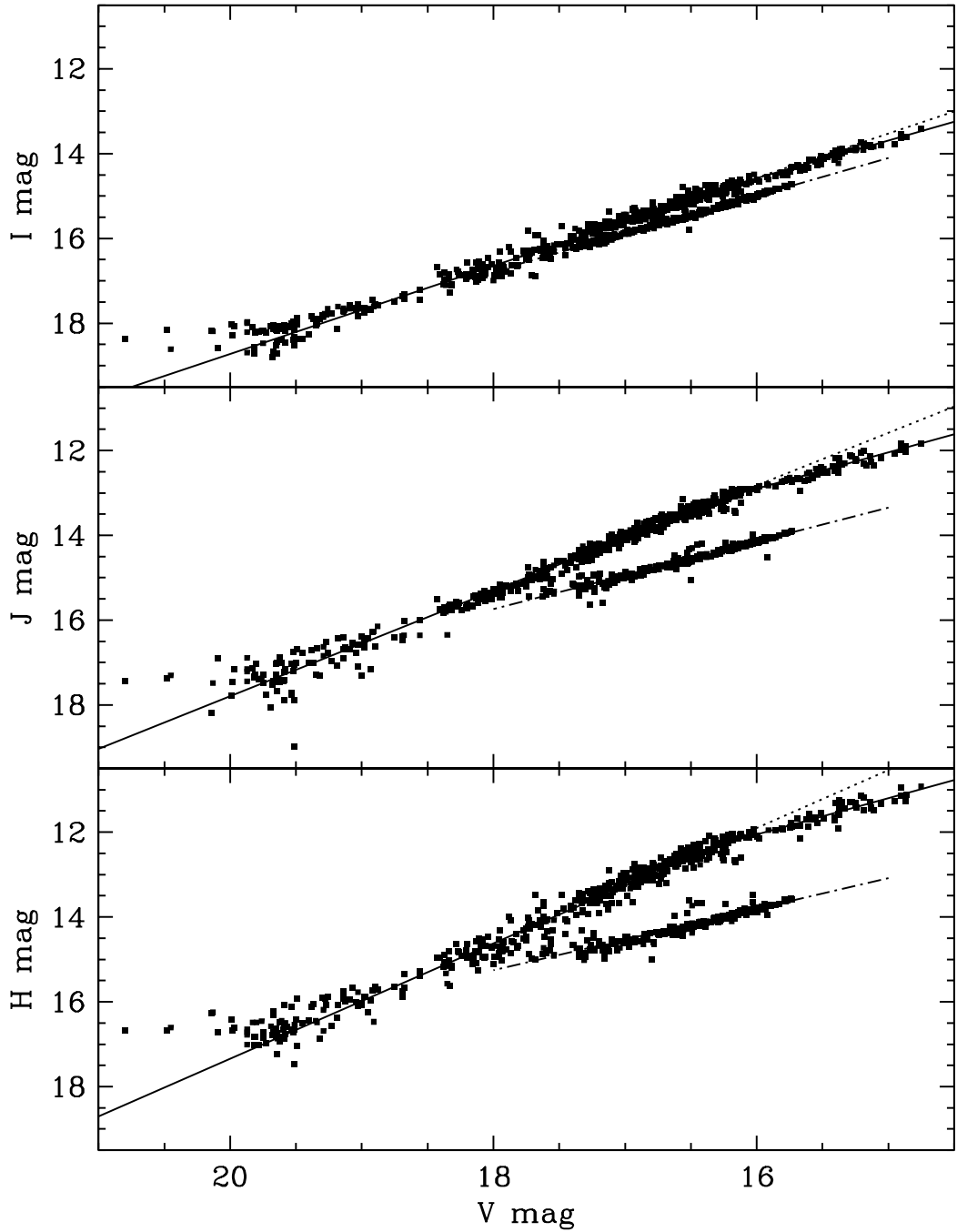}
\caption{Correlations between the $V$-band and (top) $I$-, (middle) $J$-, and (bottom) $H$-band magnitudes. The solid line is a broken power-law fit to the upper-branch; the dotted line shows an extension of the power-law before the break highlighting the difference between a single and broken power-law fit; the dot-dashed line is a single power-law fit to the lower branch data.  Best-fit parameters are given in Table \ref{corfit}.}
\label{oircorr}
\end{figure}

In Figure \ref{oircorr} we plot the $V$-band magnitudes against $I$-, $J$-, and $H$-band magnitudes.  In each plot there are clearly two separate branches, and the separation between the two branches increase from the $I$ to $H$ band. The slope of the upper branch seems to change near the brighter end.  To confirm this, we performed both single and broken power-law fits to each branch using the Levenberg-Marquardt algorithm.  The best-fit parameter values are summarized in Table \ref{corfit}.  We found the lower branch was best described by a single power-law with a slope less than one, that is, GX 339-4 became bluer as the system became brighter.  The slope of the lower branch decreased from the $I$ to $H$ band.  The upper branch, on the other hand, was best fit by a broken power-law.  From the faintest magnitudes to the break, the slope is greater than one indicating that the system gets redder with increasing brightness.  This slope increases from the $I$ to $H$ bands.  Beyond the break, the power-law slope is less than one, as in the case of the lower branch but with a higher slope in the $J$ and $H$ bands than that observed for the lower branch.  The location of the break is statistically indistinguishable (within 3-$\sigma$) between the $V/J$ and $V/H$ correlations, but is located at a slightly fainter magnitude in the $V/I$ correlation.  

The color vs. brightness trends suggest that, from the faintest optical state up to the break, the upper branches represent a state in which the O/IR is dominated by changes in non-thermal emission while at magnitudes brighter than the break, and in the lower branch, suggest changes in a thermal component is more dominant.  Since it is regarded that non-thermal emission is from collimated jets, the O/IR flux during the hard state must contain components associated directly with the inner accretion flow in addition to whatever flux is produced directly by the outer accretion disk and the (much fainter) companion star.  \cite{mai08} and \cite{rus11} presented analysis of the O/IR color-magnitude diagram for Aql X-1 and XTE J1550-564, respectively, and showed that the data could be well explained by a heated blackbody (representing the accretion disk) plus non-thermal jet emission during the hard state.  These results support the above interpretation of an accretion disk dominating the soft state, and jet emission dominating the hard state.  In a future paper we will present similar analysis of color magnitude diagrams for GX 339-4.  

\begin{figure}
\epsscale{.90}
\plotone{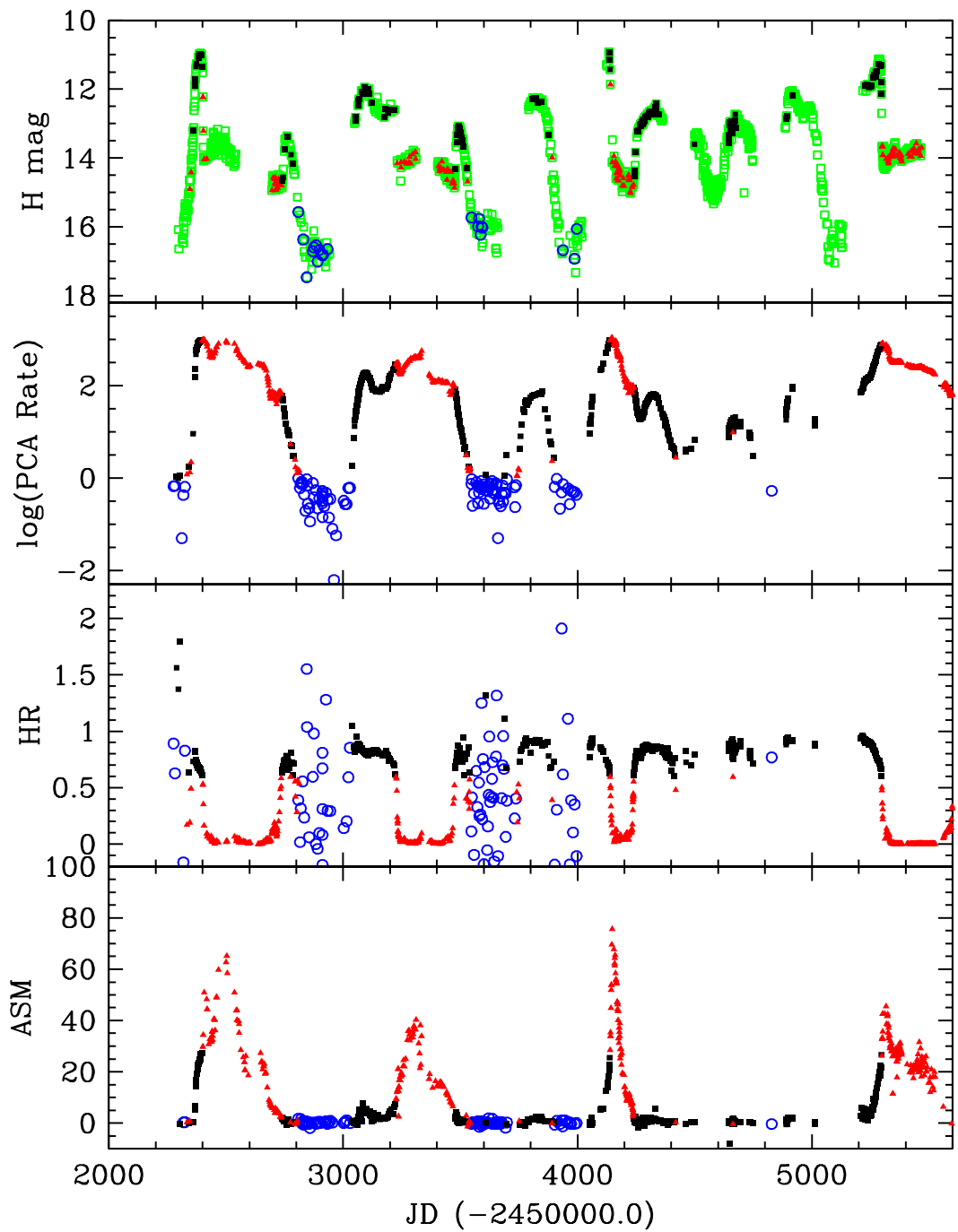}
\caption{$H$-band light curve ($top$) and RXTE data.  We categorized hard/soft state $H$-band data only for days when we had simultaneous (i.e. same day) RXTE data, else the data is shown as green open squares.  Black solid squares correspond to days when the hardness ratio is $\ge$ 0.6 (hard state) and red solid triangles when the hardness ratio is $<$ 0.6 (soft state).  Data for which the PCA count rate is $<$ 1.0 count/sec was not cateogrized and is shown as blue open circles. }
\label{oirxray}
\end{figure}

\begin{figure}
\plotone{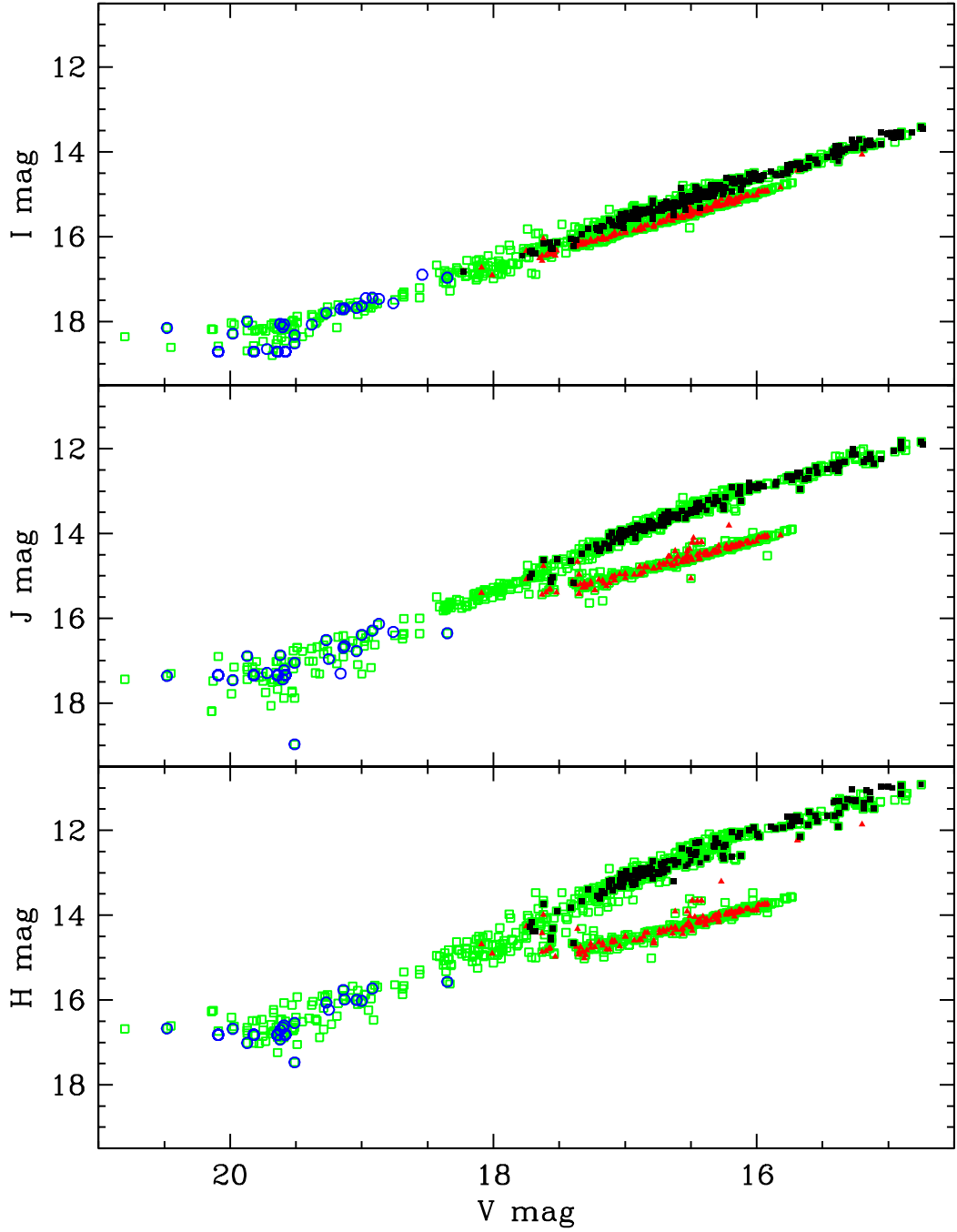}
\caption{Data are represented using the same criterion as Figure \ref{oirxray} and categorized into hard and soft states only for days when there were simultaneous RXTE data.}
\label{branchcolor}
\end{figure}

Association between the correlation slopes and the jet/disk emission could already be inferred from the results of \citet{cor09}, who explored the relationships
between O/IR flux and X-ray flux and ascertained that the upper branch corresponds to the ``hard''
X-ray state, and the lower branch to the ``soft'' state.  We confirm the association between the O/IR upper and lower branches and the X-ray
hard and soft state as follows.  First, we defined the ``soft" and ``hard" X-ray states
using the RXTE PCA hardness ratio (HR) data (where HR is the ratio of fluxes in the bands
(9.8-25 keV)/(2.8-5.2 keV)). The entire GX 339-4 RXTE dataset was analyzed to obtain count rates 
and hardness ratios. The beginning and end of outbursts in the hard state were determined 
using the light curve and setting a threshold of 10 cts/s/PCU. Then we created spectra and 
power density spectra in the 3-25 keV band for all observations with good time intervals exceeding 
1 ks while the count rate increased. We stopped the analysis when significant changes were observed 
either in the X-ray spectra or the power spectra indicating a transition to an intermediate state. 
For all these observations, the photon index was less than 1.75, and the power spectrum was 
dominated by broad Lorentzians with mean rms amplitudes greater than 25\%. 
The hardness ratio threshold including all observations was 0.6, however, for some outbursts the 
threshold can go up to 0.75.  When categorizing the O/IR into soft and hard states, 
we defined the soft state to be when HR $<$ 0.6, and the hard state when HR $\ge$ 0.6.  Second, we considered data only for days when 
simultaneous $V$- and $H$-band data were taken, as this pair of data show the separate branches most clearly.  Using these criteria, we
plotted in Figures \ref{oirxray} and \ref{branchcolor} red data points when the HR $<$ 0.6 (soft state), black points when 
HR $\ge$ 0.6 (hard state), and blue points when the X-ray state could not be determined due to low X-ray counts. We show all data as green points in Figures \ref{oirxray} and \ref{branchcolor} for completeness.
The analysis clearly shows 
that the lower (red) branch corresponds to when GX 339-4 is in a soft X-ray state, while the 
upper (black) branch corresponds to the hard X-ray state.  This confirms that we are able to use O/IR data alone to determine when GX 339-4 is in either the hard or soft X-ray state.  

\subsection{State Transitions}

\cite{cor09} discussed how GX 339-4 moved along the IR/X-ray correlation branches over the course of four separate outbursts (2002/2003, 2004/2005, 2006 and 2007).  In the following sections we trace the paths of the state transitions in the O/IR correlations and show the hard-to-soft transition in 2010.  We give a timescale over which the hard-to-soft transition occurs in 2002 and 2010.

\subsubsection{Hard-to-Soft Transition}

\begin{figure}
\plottwo{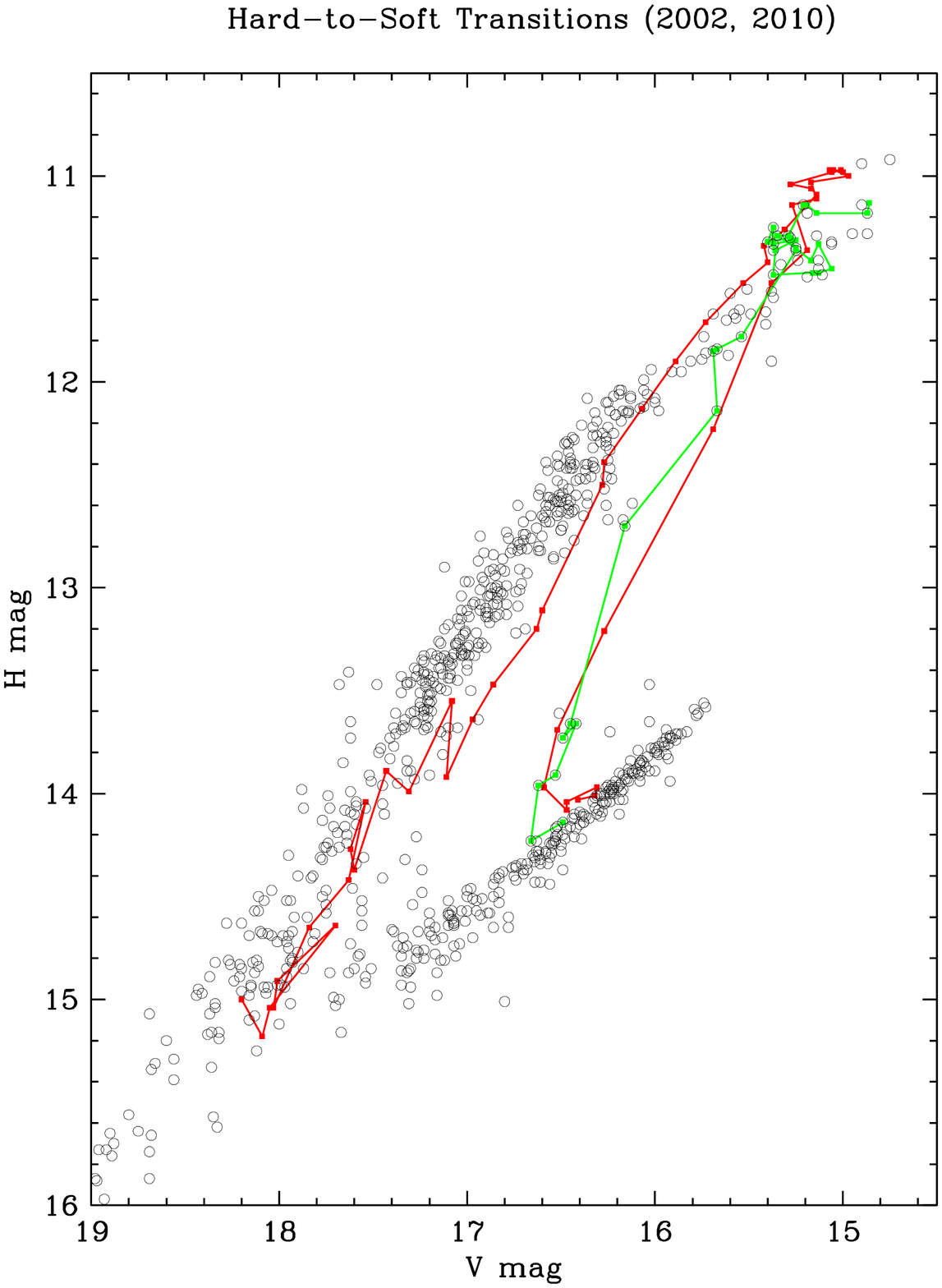}{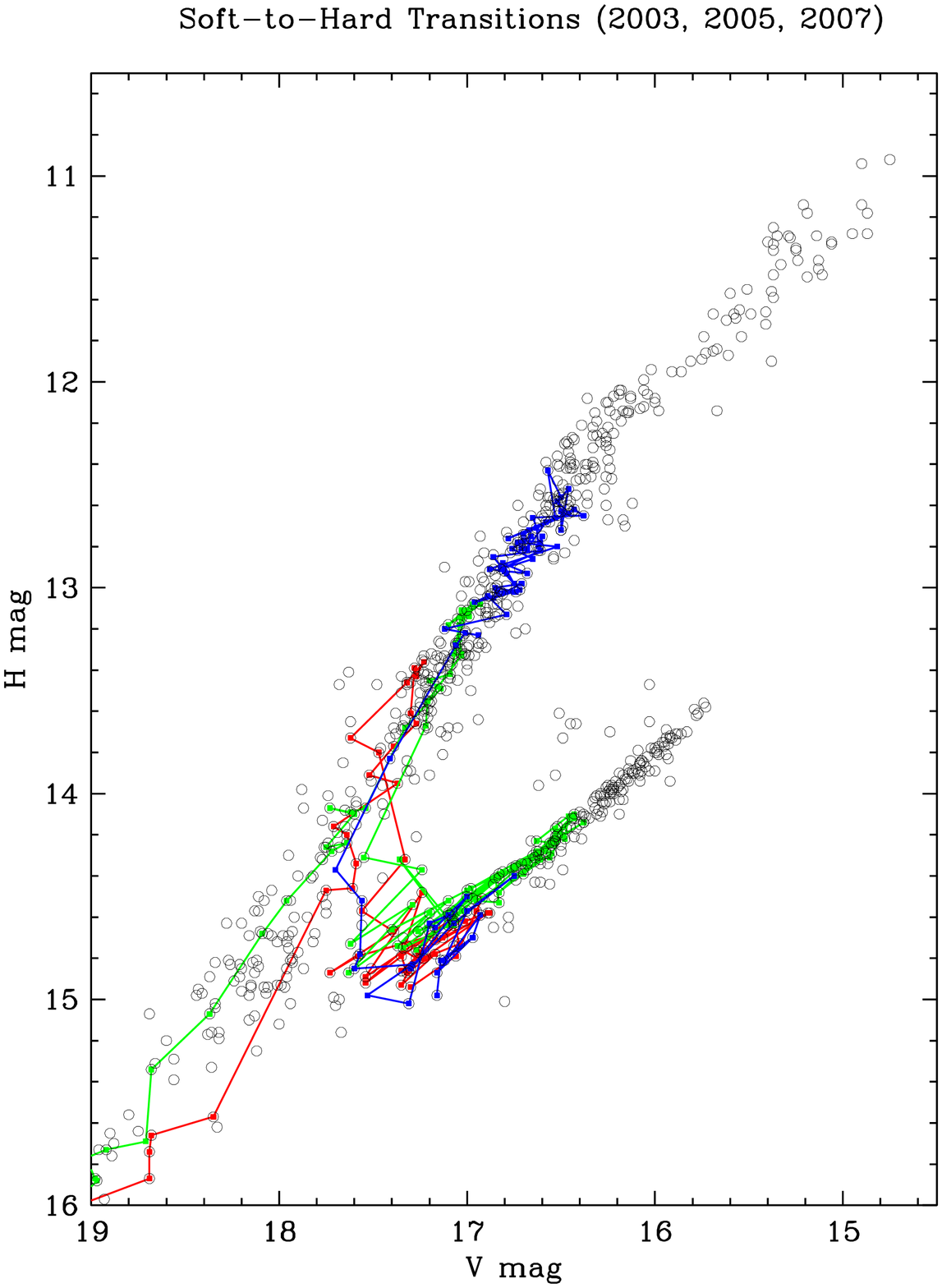}
\caption{Evolution of soft-to-hard and hard-to-soft state transitions compared to the $V/H$ correlation (open circles).  $Left:$ Hard-to-soft transition:  Red line is 2002 data, green line is 2010 data.  The soft-to-hard transition traces a unique path relative to the hard-to-soft transition.  $Right$: Soft-to-hard transition: Red line is 2003 data, green line is 2005 data, and blue is 2007. In 2003 and 2005, GX 339-4 fades to the faintest end of the branch, but remains in the hard-state branch in 2007 (and 2008/2009).  }
\label{trace}
\end{figure}

In Figure \ref{trace} (left) we show the two best examples of transitions from the hard to soft X-ray states during 2002 (red line) and 2010 (green line).  These are the two outbursts during when we had the best sampling over the hard-to-soft transition (i.e. at least one observation per day with no gaps between days over the transition period).  When GX 339-4 begins a new outburst (that is from the faintest optical state) it is in the hard state at the lower end of the hard-state branch and rises up this branch to the brightest end. This corresponds to the far right vertical branch in the HID (Figure \ref{hid}), also known as the low-hard state \cite{bel10}.  Some time after GX 339-4 reaches its peak brightness it begins to transition to the soft state (the upper horizontal branch in the HID = hard-intermediate state).  During this phase GX 339-4 rapidly fades, more in the IR than the optical. We note that \cite{mot85} was the first to see a change in optical brightness over the transition from hard to soft states. This sudden drop in O/IR flux during 2002 was also noted by \cite{hom05}.  In both 2002 and 2010 the path taken by GX 339-4 over this transition is remarkably similar.  Once it has left the hard-state branch it takes around 4 days to reach the soft state branch.  Both cases merge with the soft branch at around the same location: $V \sim$ 16.5 mag. From our data, and as stated in \cite{cor09}, we can only say that the rise of GX 339-4 follows the same path as the decay phase (see Section \ref{shtrans}) from a $V$-band magnitude of around 18.2.  We do not have any data to state definitively that the rise and decay paths are the same below this magnitude. 

\subsubsection{Soft-to-Hard Transition}
\label{shtrans}

When GX 339-4 is in the soft X-ray state (the left vertical branch in the HID = high-soft state) it is already at relatively bright O/IR fluxes.  In Figure \ref{trace} (right) we show three examples (2003, 2005, 2007) of a soft-to-hard transition (corresponding to the bottom horizontal branch in the HID).  During 2003 and 2005 the transition from soft to hard X-ray states followed the same path: GX 339-4 moves down the soft state branch until it reaches the intersection between the soft and hard state branches.  It then moves up the hard state branch to a peak magnitude (slightly brighter in 2005 than 2003 by around 0.2 mag in $V$) and then swiftly down the hard state branch to the faintest end.  After transitioning into the soft state in 2007 GX 339-4 did not move down to the faintest end of the hard state branch until 2009.  Throughout this period GX 339-4 moved up and down the middle portion of the hard state branch.  Toward the end of 2009 it finally moved down the hard state branch to the faintest end.  

\section{Spectral Energy Distributions}\label{seds}

\subsection{Determination of O/IR Flux}\label{fluxcal}

To obtain fluxes appropriate for comparing with observations in other wavelength
regimes requires two steps.  First, the magnitudes must be dereddened so that they
are not affected by absorption along the line of sight; second, the dereddened 
magnitudes must be converted to flux units.

The O/IR magnitudes were dereddened using E(B-V) = 1.2 $\pm$ 0.1 \citep{zdz98} that was converted to A$_V$ = 3.7 $\pm$ 0.3 \citep{car89}.  We calculated the extinction coefficient in the other bands using $A(\lambda)/A(V) = a(x) + b(x)/R_{V}$ and the relationships for $a(x)$ and $b(x)$ as given in \cite{odo94} for $V$ and $I$ bands, and \cite{car89} for $J$ and $H$ bands.  In calculating $a(x)$ and $b(x)$, we used the effective wavelength for each band \citep{fro78,eli82,bes98}: $V$ = 545 nm, $I$ = 798 nm, $J$ = 1250 nm, $H$ = 1650 nm.
The errors on the dereddened magnitudes are dominated by
the error in A$_V$, and are as follows:  $V_{err}$ = 0.3 mag, $I_{err}$ = 0.1 mag, 
$J_{err}$ = 0.1 mag, $H_{err}$ = 0.1 mag.  Adding the photometric and interstellar redenning errors in quadrature give the following total errors on dereddened magnitudes: $V_{err}$ = 0.30 mag, $I_{err}$ = 0.10 mag, $J_{err}$ = 0.10 mag, $H_{err}$ = 0.10 mag.

We note that \cite{hom05} found an independent value of E(B-V) using $N_H = 5 \pm 1 \times 10^{21}$ atoms cm$^{-2}$ \citep{kon00}, converting to E(B-V) = 0.94 $\pm$ 0.19 \citep{pre95}, in turn giving $A_V$ = 2.9 $\pm$ 0.6 \citep{car89}.  \cite{zdz98} obtained their value of E(B-V) from a detailed study of extinction of $\sim$450 stars within $\pm 5^o$ of GX 339-4.  In either case, E(B-V) agree within the errors.  

To convert the dereddened magnitudes to flux (in units of Jy) we used the following zeropoint fluxes (where the zeropoint flux of a given filter is that corresponding to zero magnitude):  
$V_0$ = 3636 Jy, $I_0$ = 2416 Jy \citep{bes98}, $J_0$ = 1670 Jy, $H_0$ = 980 Jy \citep{fro78,eli82}.  The dereddened fluxes of GX 339-4 for each waveband are given in Table \ref{alldata}.

\begin{figure}
\plottwo{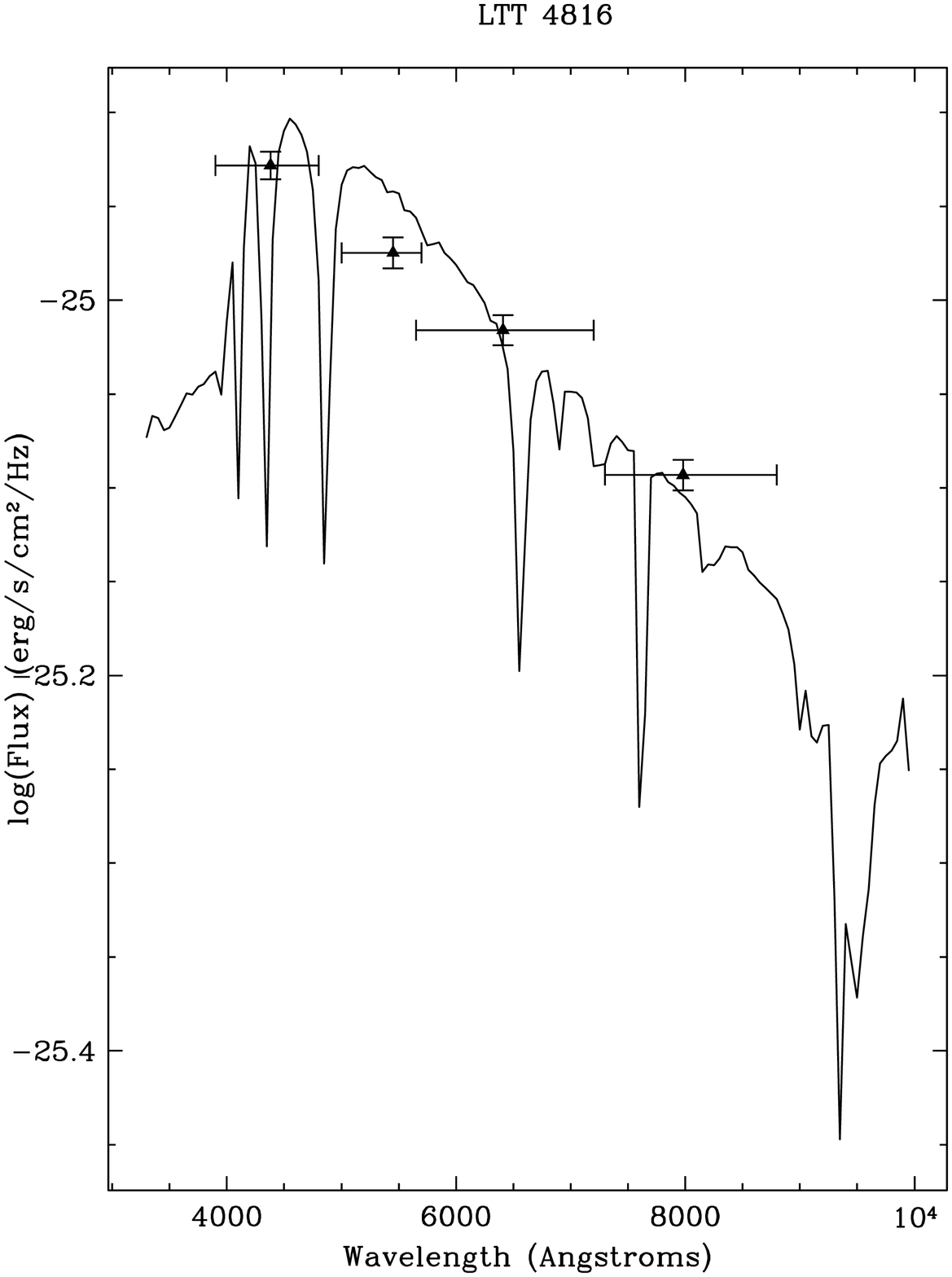}{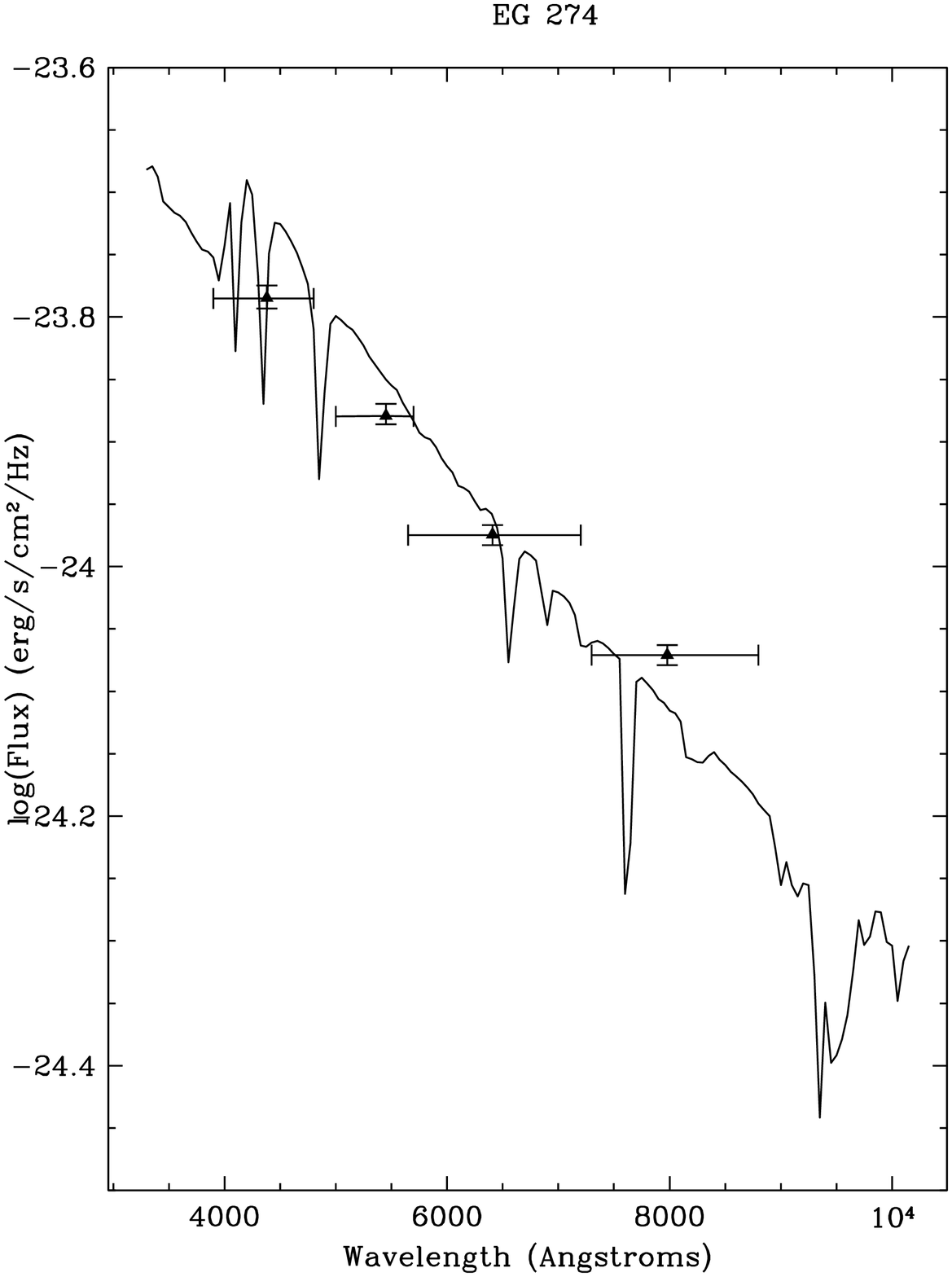}
\caption{Spectrophotometric standards ($left$) LTT 4816 and ($right$) EG 274.  Triangles are, from left to right, SMARTS $B$, $V$, $R$, and $I$ data calibrated using methods outlined in Sections \ref{magcal} and \ref{fluxcal}.  Vertical errorbars represent the 0.02 mag calibration error, and horizontal errorbars show the full-width-half-maximum of the filter bandpass.}
\label{specphot}
\end{figure}

We have checked our flux calibration method by obtaining observations of optical 
spectrophotometric standards (LTT 3218, LTT 4816, EG 131, EG 274) and comparing our fluxes to those observed by \cite{ham92}, \cite{ham94}, and \cite{bes99}.  
Observations were obtained in $B, V, R$, and $I$-bands and our magnitudes were converted into fluxes using the 
method outlined in Section \ref{magcal}.  We compared the photometric fluxes to the flux calibrated spectrum over the relevant bandpass, and found no deviations larger than 
0.05 dex, or 10\% of a given flux.  We show two examples (LTT 4816 and EG 274) in Figure \ref{specphot}.

\subsection{Soft state SED}\label{softsed}

\begin{figure}
\epsscale{.80}
\plotone{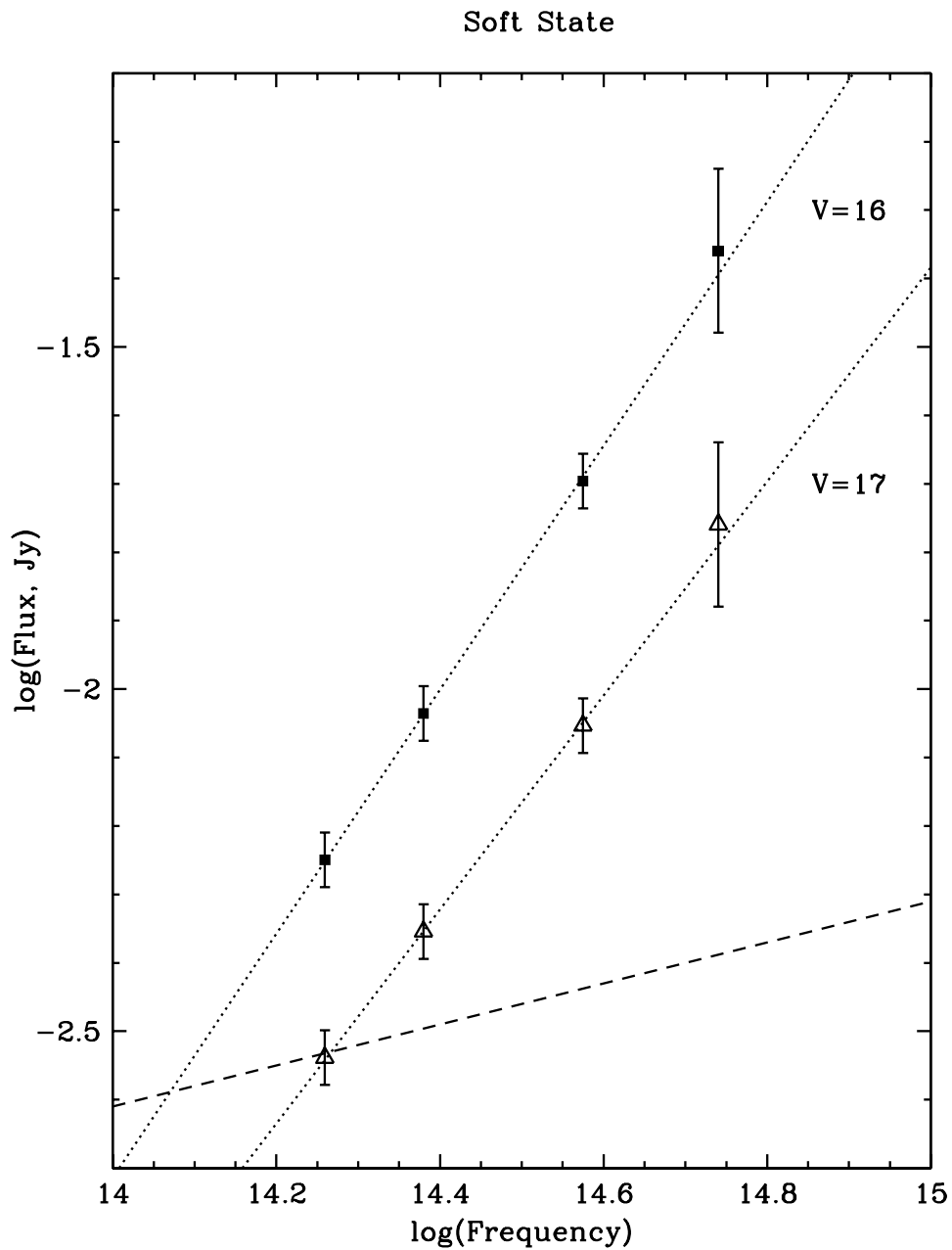}
\caption{Soft-state SEDs of GX 339-4 for V = 16 and 17 mag.  $I-, J$-, and $H$-band data were extrapolated from the power-law fits to the correlations shown in Figure \ref{oircorr}.  The errors are the combined photometric and reddening errors as outlined in Section \ref{fluxcal}. The dashed line represents the spectral slope expected for a viscous-heated accretion disk ($F \propto \nu^{1/3}$).  Dotted lines are the best-fit power-laws to the OIR data; $\alpha_{v=16} = 1.8 \pm 0.1$; $\alpha_{v=17} = 1.6 \pm 0.1$. }
\label{softsedv16}
\end{figure}

As stated in Section \ref{oircorrs}, our fits to the slope of the soft-state branch indicate that the flux increases 
more quickly at bluer wavelengths.  In other words, the source gets bluer as it gets brighter. 
Figure \ref{softsedv16} shows two example SEDs obtained from the soft-state branch 
data for $V =$ 16.0 and 17.0 mag.  The $J$, $H$ and $I$-band data were extrapolated from the power-law fit to the correlation data shown in Figure \ref{oircorr}.  In both cases, the O/IR data lie on the same spectral slope.  We determined the slope by fitting a 1-deg polynomial in log-log space using the Levenberg-Marquardt method and derived the following results:  $\alpha = 1.6 \pm 0.1$ for $V$ = 17 mag; $\alpha = 1.8 \pm 0.1$ for $V$ = 16 mag.  Hence, the slopes in each case are the same, within errors.  We show these fits as dotted lines in Figure \ref{softsedv16}.  Also shown, as a dashed line, is the slope expected for $F \propto \nu^{1/3}$ representing a viscously-heated steady-state accretion disk.  The O/IR SEDs during the soft state have spectral slopes closer to that expected for the RJ tail ($\alpha = 2$).  An RJ tail may be present from either the outer regions of a viscous-heated disk or an irradiated disk.  Since we do not observe a spectral turnover we cannot say definitively what the temperature of the thermal source is nor what the heating mechanism is.  In any case, the trend seen during the soft state in GX 339-4 with higher fluxes is consistent with a thermal source that is heated to higher temperatures.

\subsection{Hard state SED}\label{hardsed}

In Section \ref{oircorrs} we found that the hard state branch was best described by a broken power-law where at fainter magnitudes GX 339-4 gets redder at higher fluxes while beyond the break the opposite occurs.  For the former case, this requires either a non-thermal source or a thermal source that dramatically increases in size as it decreases in temperature.  For the latter case, GX 339-4 is getting hotter with higher fluxes.  

\begin{figure}
\epsscale{.80}
\plotone{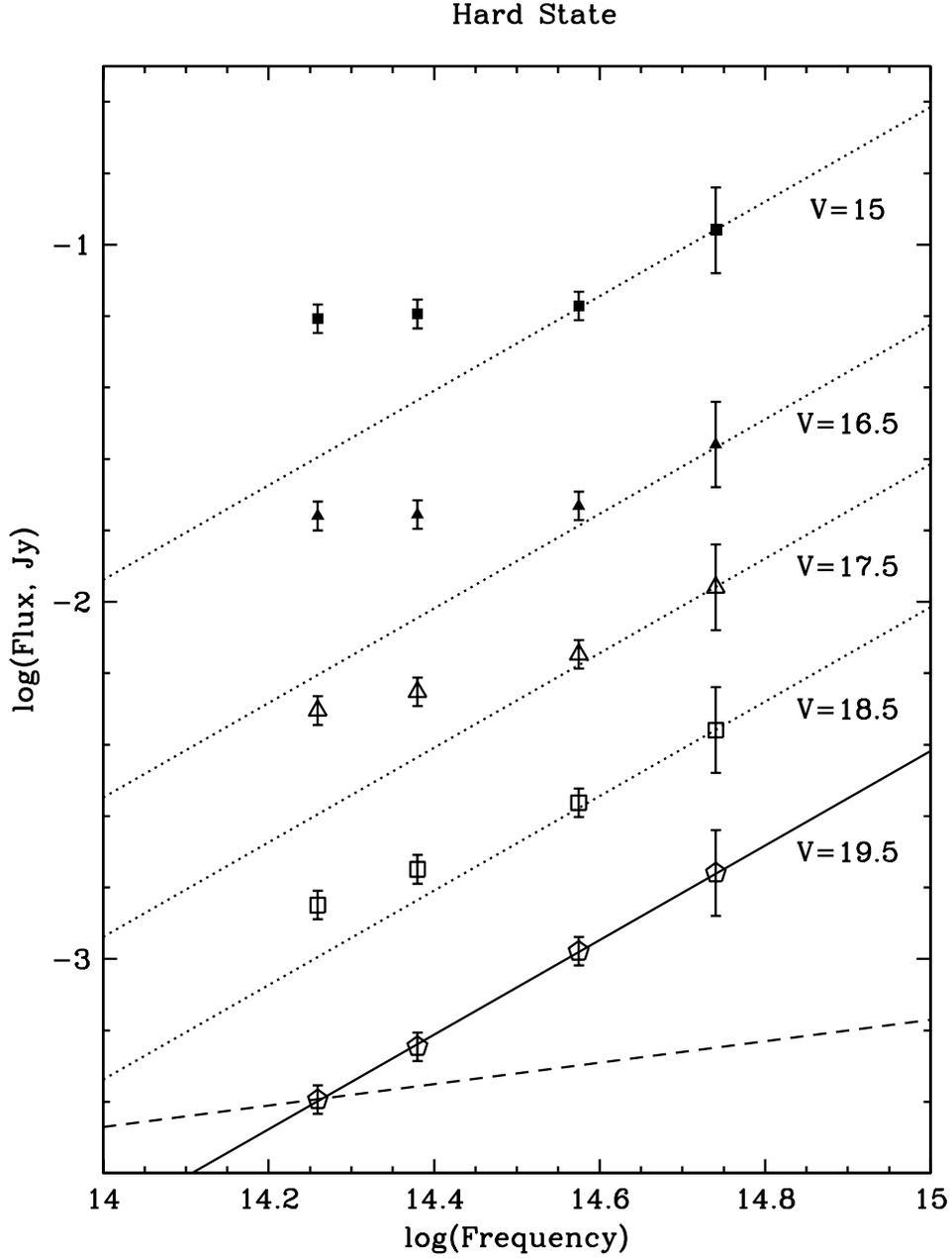}
\caption{Hard state SEDs of GX 339-4 for $V$ = 19.5 - 15.0 mag.  Dashed line represents a viscous-heated accretion disk ($F \propto \nu^{1/3}$).  The solid line is a best-fit power-law to the O/IR data ($\alpha = 1.3 \pm 0.1$) at $V$ = 19.5 mag.  This line is repeated as dotted lines and shifted up to match the $V$-band, highlighting that the $V$ and $I$ bands remain on the same spectral slope as GX 339-4 increases in flux.  The $J$ and $H$ band spectral slope decreases with increasing brightness but the near-IR flux also increases suggesting an additional, non-thermal source becomes more dominant as GX 339-4 increases in flux.}
\label{hardsed1}
\end{figure}

Figure \ref{hardsed1} shows the evolution of the hard-state O/IR SED from the faintest $V$-band magnitude ($V = 19.5$) to the brightest observed on the hard-state branch ($V = 15$).  At the faintest magnitude, the O/IR data lie on the same spectral slope ($\alpha = 1.3 \pm 0.1$).  This is shown as the solid line in Figure \ref{hardsed1}.  This slope is less than the slope observed during the soft state.  The spectral slope expected for a viscous-heated accretion disk ($\alpha \sim 1/3$) is also shown as the dashed line.  As for the soft state, we see that the optical bands have a much steeper slope but we are unable to discern the temperature or heating mechanism of the thermal source.  Nevertheless, even when GX 339-4 is at its faintest optical state we see evidence of a heated thermal source, most likely the accretion disk.  In Figure \ref{hardsed1} we have propagated the fitted slope at the faintest magnitude to the SEDs at brighter magnitudes to show that, as GX 339-4 increases in brightness, the $V$ and $I$ bands are consistent with this slope, indicating that $V$ and $I$ are most likely dominated by a thermal source that is heated to higher temperatures with increasing O/IR flux.  \cite{hom05} also reported a stable spectral slope between the $V$ and $I$ bands during the hard state rise and transition to the soft state during the 2002/03 outburst.

The near-IR bands, on the other hand, show a clear evolution in slope that decreases with increasing flux, while the flux in the near-IR bands also increases.  This was also observed by \cite{hom05} during the 2002/03 outburst.  

Radio emission in GX 339-4 and other XRBs has only been observed during the hard state and transitions from one state to another \citep{fen01,cor02,cor03,fen04,gal04,fen09}.  Presumably the radio emission is due to synchrotron emission from a jet.  It has been suggested that such emission might contribute significantly at near-IR bandpasses \citep{cor02,now05} and explain the near-infrared excess we see in our hard-state SED.  We searched the literature for radio observations that were taken on the same night as the SMARTS data during the hard state.  We found two cases.  The first was taken on UT 2002 April 03 \citep{now05} with ATCA at 1.38 GHz (4.83 $\pm$ 0.20 mJy) and 2.40 GHz (5.13 $\pm$ 0.11 mJy).  On this date, we observed GX 339-4 with a $V$-band magnitude of 15.73.  The second case was on UT 2002 April 18 \citep{now05} at 8640 MHz (13.49 $\pm$ 0.08 mJy) and 4800 MHz (12.97 $\pm$ 0.07 mJy).  Note that this data was first reported in \cite{gal04} but was re-analyzed by \cite{now05}.  On this date, GX 339-4 had a $V$-band magnitude of 15.14 mag.  We show these radio data along with our O/IR hard-state SEDs for these $V$-band magnitudes in Figure \ref{hardsed2}.  We performed power-law fits to 1) the radio, $J$ and $H$ band data; and 2) radio data only, using a deg-1 polynomial and the Levenberg-Marquardt method.  These fits are shown in Figure \ref{hardsed2} as the solid and dotted lines, respectively.  The best-fit slopes are for SED 1 (03 April): $\alpha_{radio/IR}$ = 0.14 $\pm$ 0.01; $\alpha_{radio} = 0.11 \pm 0.09$; and for SED 2 (18 April): $\alpha_{radio/IR}$ = 0.14 $\pm$ 0.01; $\alpha_{radio} = 0.14 \pm 0.01$.

On UT 03 April 2002, the error on the radio data is sufficiently large that a simple extrapolation from the radio to the near-IR spectral may or may not be consistent with the O/IR data.  A fit to the radio and near-IR data on this date has the same spectral slope as that found on UT 2002 April 18.  The latter has much tighter constraints thanks to the relatively small errors on the radio data.  Indeed, the spectral slopes are the same with and without the near-IR data included in the fit.  Interestingly, the $I$-band also lies on this extrapolation.  Since the $I$-band lies on the same spectral slope as the $V$-band from the faintest to brightest magnitudes during the hard-state (i.e. the $V-I$ color is constant) then the $I$ band is most likely the location of a break between the non-thermal and thermal SEDs.  We do not observe a break in the non-thermal spectrum, however, we can say that the non-thermal break location must exist at frequencies greater than or equal to the $J$-band, in agreement with \cite{cor09} but contrary to \cite{gan11} (see Section \ref{discuss} for further discussion of this point).  

At magnitudes brighter than the break in the hard-state branch, GX 339-4 gets bluer at higher fluxes.  The SED shown in Figure \ref{hardsed1} shows that the non-thermal component is present, even at the brightest magnitudes (i.e. beyond the break).  However, the near-IR flux at $V$ = 15.0 mag relative to the $V$-$I$ spectral slope is less than at $V$ = 16.5 (just before the break).  In other words, as the $V$-$I$ fluxes increase beyond this break, the near-IR fluxes increase, but less so than at magnitudes below the break. Clearly either a physical change has occurred in the jet and/or the jet output may be reaching a saturation point as GX 339-4 gets brighter. \cite{cor09} suggest that the break in the jet spectrum SED moves to higher frequencies at higher fluxes.  Below the break in the hard-state branch we may be observing optically thin jet emission while above the break at higher fluxes we may be observing optically thick jet emission.  We leave further investigation of this point to a future paper.  In addition, changes in the thermal emission become more dominant beyond the break.  Since the $V-I$ slope is the same this simply translates to the thermal source (i.e. disk) getting hotter with increasing flux.  

In summary, simultaneous radio and near-IR data can be explained by a non-thermal component, and this component dominates at all but the faintest magnitudes.  The optical spectral region remains dominated by a thermal flux source, most likely a heated accretion disk.  Therefore, the O/IR SED needs to be explained by a combination of thermal and power-law models.  We are unable to perform such a fit since the number of parameters involved would exceed the number of constraints (data) available.  This highlights the need to obtain simultaneous data at other wavelengths, such as $B$, UV, and mid-IR in addition to the O/IR, radio and X-ray, to further sample the shape of the SED near the O/IR and to best constrain model fits. In addition, obtaining simultaneous radio data with small errors and/or as many bandpasses as possible would strengthen model fits and place more confidence in associating near-IR flux with jet emission.

\section{Timing}\label{timing}

We have calculated the power spectral density (PSD) of the variability of GX 339-4 
in the hard and soft states. The PSD corresponds to the power in the variability of emission 
as a function of timescale. The power in the units of rms$^2$ Hz$^{-1}$ is given by:
P($\nu$)=A$\nu^{\alpha}$,
where A is a normalization constant, $\nu$ is the frequency (inverse of timescale) in Hz and $\alpha$ is the power law index. In a log[P($\nu$)] vs. log[$\nu$] plot, $\alpha$ is defined as the slope of the power-law power spectrum.  
The timescales covered in the power spectra range from $\sim$8 months (the longest well-sampled intervals during which any state transition did not take place) to 1--2 days (the most frequent sampling carried out for a significant interval). Hence, the power spectra range from $10^{-7.3}$ Hz to $10^{-5.3}$ Hz. 

We divided the $V$- and $H$-band light curves into ``soft" and ``hard" intervals, respectively, 
based on the hardness ratio in the contemporaneous RXTE-PCA observations, as outlined in Section \ref{oirstates}. These two bands were chosen as the $H$-band data shows light-curve variability most clearly during the hard state, while the same is true for the $V$-band during the soft state.  However, PSDs were calculated for each band in both the hard and soft states.  The time intervals chosen were $V$: JD = 2453224.70 - 2453478.79; and $H$: JD = 2454498.88 - 2454746.55.  These were the longest well-sampled intervals during which the X-ray hardness ratio stayed soft and hard, respectively.  Over these time intervals, the $V$-band magnitude ranged between 15.9 and 17.6 mags, while the $H$-band varied between 12.7 and 15.3 mags.  

The raw PSD calculated from a light curve combines two aspects of the dataset: (1) the intrinsic variation of the object and (2) the effects of the temporal sampling pattern of the observations. In order to remove the latter, we applied a Monte-Carlo type algorithm based on the ``Power Spectrum Response Method'' (PSRESP) of \citet{utt02} to determine the intrinsic PSD (and its associated uncertainties) of the light curves. Our realization of PSRESP is described in \citet{cha08}. PSRESP gives both the best-fit PSD model and a ``success fraction'' $F_{\rm succ}$ (fraction of simulated light curves that successfully represent the observed light curve) that indicates the goodness of fit of the model. We bin the data in $1$ to $2$-day time intervals, averaging all data points within each bin to calculate the flux. We filled empty bins through linear interpolation of the adjacent bins in order to avoid gaps that would distort the PSD. We accounted for the effects of the binning and interpolation by inserting in each of the simulated light curves the same gaps as occur in the actual data and performing the same binning and interpolation procedures.

\begin{figure}
\epsscale{.80}
\plotone{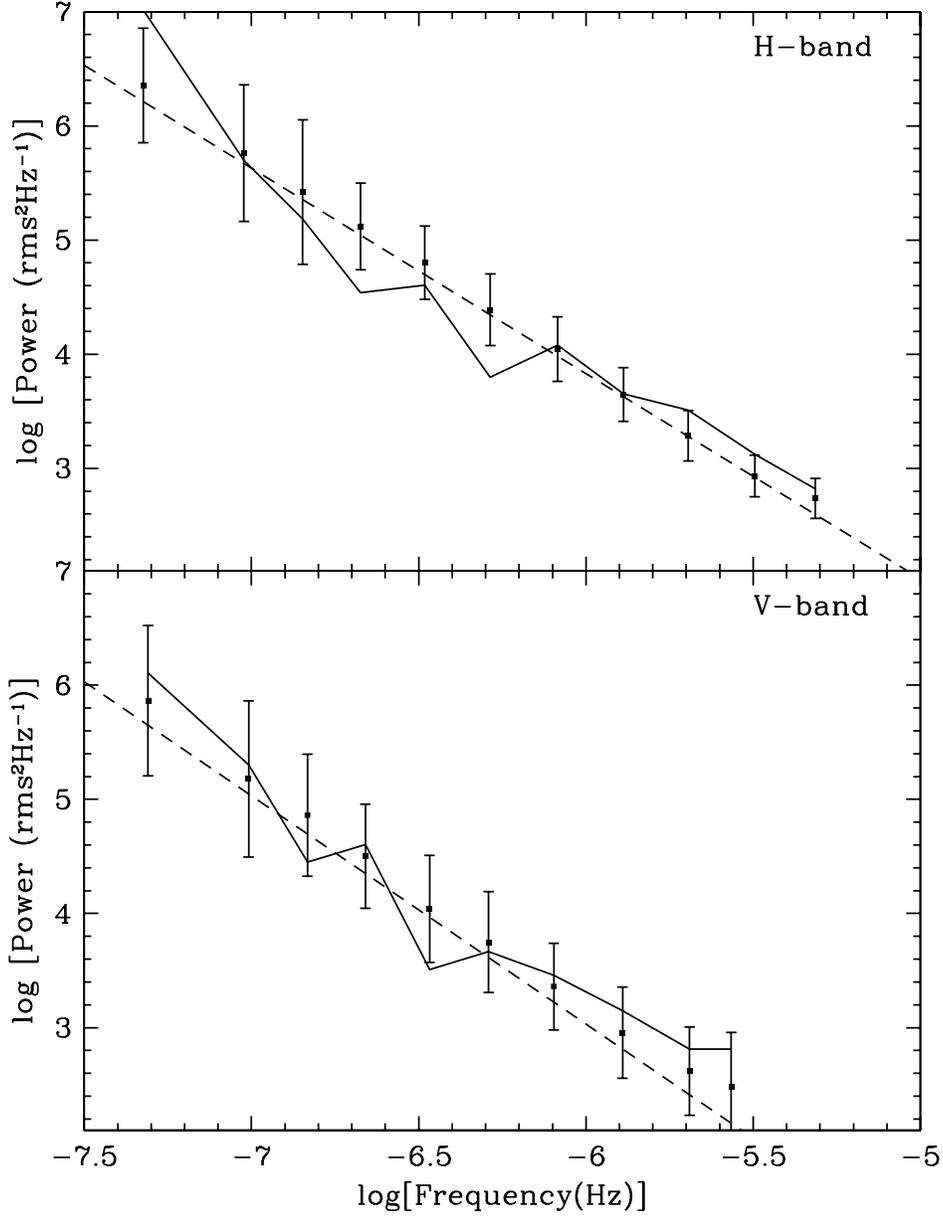}
\caption{Power spectral densities of the hard state $H$-band, and soft state $V$-band, light curves. The PSD of the observed data is shown by the solid jagged line, while the underlying power-law model is shown by the dashed straight line. Points with error bars correspond to the mean value of the PSD simulated from the underlying power-law model. The error bars are the standard deviation of the distribution of simulated PSDs. The broadband power spectral density is best described by a simple power law with a slope $-1.8$ at $H$-band, and $-2.0$ at $V$-band.  See Table \ref{psdfits} for the PSD fit results for the $V$-band during the hard state and $H$-band during the soft state.}
\label{psd}
\end{figure}

In Figure \ref{psd} we show the PSDs of GX 339-4 and the best-fit model in $H$ (hard state) and $V$ bands (soft state).  The PSDs show red noise behavior, i.e., there is higher amplitude variability on longer than on shorter timescales. Based on the model with the highest success fraction, the $H$-band PSD, and therefore the hard-state PSD, is best fit with a simple power law of slope $-1.8^{+0.3}_{-0.7}$, for which the success fraction is $0.56$. That for the V-band (soft state) PSD is $-2.0^{+0.3}_{-0.5}$ with success fraction $0.89$. During the fitting, we varied the slope from $-1.0$ to $-3.0$ in steps of 0.1. The uncertainties of the slopes represent the range of slopes beyond which the success fraction goes below half of the maximum value. The rejection confidence, equal to one minus the success fraction, is much less than 0.9 in both cases. This implies that a simple power-law model provides an acceptable fit to the PSD at both wavebands.  The PSD for the $V$ band during the hard state, and $H$ band during the soft state, were also well-fitted by a simple power law model with slopes consistent with those described here within uncertainties. A summary of our PSD fits are given in Table \ref{psdfits}.  We see no break in the power spectra, nor evidence of any QPO's, at few days to few months timescales. 

The standard deviations of the light curves are 7.8 mJy ($V$-band) and 3.6 mJy ($H$-band). The mean fluxes are 27.6 mJy ($V$-band) and 6.2 mJy ($H$-band). The uncertainties are negligible compared to the variation.  To meaningfully compare the fluctuation of two light curves, we need to i) normalize the standard deviation by the mean of the light curves ii) account for the contribution of the observational uncertainties, as opposed to the intrinsic fluctuations, to the observed variability of the light curves. Here, we calculated the ``excess variance" normalized by the square of the mean flux, as a measure of the variability in the $V$- and $H$-band data. It is defined as
\begin{equation}
F_{\rm var}=\sqrt{{\frac{\rm S^2-\overline{\sigma_{\rm err}^2}}{\overline{x}^2}}},
\end{equation}
where $\rm S$ is the variance, $\sigma_{\rm err}$ is the observational uncertainty, and $\overline{x}$ is the mean of the data \citep{nan97,ede02,vau03}. We find that the values of $F_{\rm var}$ of the $V$- and $H$-band variability for the entire data set are $0.28$ and $0.58$, respectively, i.e, of the same order of magnitude with the $H$-band having a higher value. This is consistent with the slightly larger amplitude of power in the $H$- compared to the $V$-band as seen in Figure \ref{psd}.

The timescales covered in Casella et al. (2010) and Gandhi et al. (2010) are from few minutes to 0.1 sec (0.003 Hz to 10 Hz). Therefore, there is a large gap between the frequency range covered in those papers and that presented here. The slope at the lowest frequency end of the power spectrum shown in those work is significantly flatter than the slope determined here for either of the wave bands. But the data used in this paper is insufficient to determine at which frequency the flattening takes place.

Well-sampled light curves covering minutes to days timescales can be used to fill-up the gap in the frequency range and check the broadband nature of the power spectrum from months to sub-second timescales. This may indicate whether the slope changes gradually over a range of timescales, or it changes abruptly and in the latter case, the numerical value of that characteristic timescale. This will be carried out in a future paper.

\section{Discussion and Conclusions}\label{discuss}

We have presented a comprehensive optical and near-infrared light curve of GX 339-4 
between 2002 and 2010, inclusive.  The light curve shows that GX 339-4 is continuously in outburst (that is, the companion star never dominates the O/IR flux) but experiences significant changes in luminosity and spectral state.  

When GX 339-4 transitions from the hard state to the soft state changes in the near-IR fluxes are much greater than in the optical fluxes.  This can be understood if the near-IR flux is dominated by non-thermal emission from, say, a jet.  The change in near-IR flux is significantly greater during the hard-to-soft transition than the soft-to-hard transition (i.e. rebrightening event) suggesting that non-thermal emission is more dominant during the hard state before the hard-to-soft transition than it is after the soft-to-hard transition.  

We have seen three examples of rebrightening events during 2003, 2005, and 2007.  However, the 2007 event was markedly different to the others in that GX 339-4 did not return to quiescence after the peak was reached, and remained in the hard state during 2008 and 2009 during which it fluctuated significantly in O/IR fluxes.  Rebrightening events have been observed in the O/IR during the outbursts of 4U1543-47 \citep{bux04} and XTE J1550-564 \citep{jai01}, and in optical/X-ray fluxes in XTE J1752-223 \citep{rus12}.  The shape of the rebrightening events in all three systems are remarkably similar in that they exhibit a fast-rise-exponential-decay profile.  The increase in O/IR occurs soon after the source transitions from the soft to the hard X-ray state \citep[via the intermediate state in 4U 1543-47,][]{kal05} along with a resurgence in radio emission \citep{cor01,jai01,kal05}.  The source then decays back to a 
magnitude where it remains steady until the next outburst (except for the 2007 event in GX 339-4).  Since: 1) these events occur after transition into the hard state; 2) the flux becomes redder with increasing flux; and 3) radio emission is detected near the peak of the O/IR maximum, it seems likely that rebrightening events are associated with the reappearance of collimated jets.  The rebrightening events observed in GX 339-4 have differences between their durations and the final magnitudes reached after the peak has faded. Only one rebrightening event has been observed in 4U 1543-47, XTE J1550-564, and XTE J1752-223 so we cannot comment on any differences that may occur from one outburst to the
next in these systems.  However, the duration of the rebrightening events in other BHXRBs are significantly different to that of GX 339-4: $\sim$ 60 days in XTE J1550-564, $\sim$ 35 days in 4U 1543-47, and $\sim$ 40 days in XTE J1752-223.  These differences may provide clues as to how much material has been either accreted by the black hole or ejected via the jets.  Research is now starting to focus on how much jets contribute to the overall flux of the system at high and low luminosities \citep[e.g.][]{rus10,sol11} and decomposition of the disk/jet fluxes \citep[e.g.][Din\c{c}er et al., submitted]{mai09,rus10,rus11,rus12}.  For GX 339-4, we are now able to observe multiple rebrightening events for one BHXRB.  The differences we have observed in the rebrightening events may provide interesting and crucial limits to theoretical models, and we may now begin to study whether these differences are correlated with other properties of the outbursts.  

The HID of GX 339-4 is broadly similar for all outbursts except in 2006 when GX 339-4 failed to travel around the HID loop.  Based on an empirical relationship between the hard X-ray peak flux and outburst waiting time, \cite{yu07} and \cite{wu10} suggest that, assuming that the hard X-ray flux produced is directly related to the amount of mass accreted from the disk, there exists a disk mass limit (as yet unknown) that must be exceeded in order for GX 339-4 to generate a hard X-ray outburst above $\sim$ 0.2 Crab.  In 2006, GX 339-4 only reached $\sim$ 0.1 Crab, below this limit.  Future studies may shed light on what this limit is physically and quantitatively, and how it is related to the HID.  GX 339-4 experienced an outburst in 2010 that was above this limit.  However, it occurred somewhat later and had a lower peak flux ($\sim$ 0.4 Crab) than that prediced by \cite{wu10}.  The other outbursts move around the loop counterclockwise, however there are some differences 1) between the upper and lower horizontal branches, especially during 2004/05; 2) the extent to which GX 339-4 reaches the far left part of the diagram; and 3) the amount of time spent in each branch of the HID.  We may now be close to a position that allows us to start studying correlations between the various properties in the HID (X-ray) and the O/IR outbursts.

We have found strong correlations within the O/IR with two branches corresponding 
to either the hard or soft X-ray states.  These correlations are similar to those seen 
between O/IR and X-ray flux \citep{cor09} with distinct branches corresponding to the 
hard and soft states.  With the correlations so clearly defined, particularly between $V$- and $H$-band, it is possible to determine whether GX 339-4 resides in the hard or soft X-ray state from the O/IR data alone.  We have found the soft state branch is best described by a single power-law while the hard-state branch is best represented by a broken power-law.  \cite{cor09} also found a break in their hard state branch, but only in the $H$-band/X-ray correlation (not in the $V$-band/X-ray correlation) and at significantly lower $H$-band fluxes than that observed in this work. Their break occurs just before the the hard-state branch bifurcates whereas we observe a break near the brightest end of the hard-state branch.  In future work, we will explore the difference in break location and how this break may or may not relate to the properties of the jet spectrum.

In this paper we showed the paths traced in the O/IR correlation space when the hard-to-soft and soft-to-hard state transitions occurred.  For the former, we showed two well-sampled examples (2002, 2010) and that this transition occurs very quickly ($\sim$4 days).  The paths taken in these two cases were very similar and ended the transition in the soft-state branch at around the same magnitude: $V \sim 16.5$ mag.  For the latter case we showed three cases.  In two (2003 and 2005) we saw very similar behavior where the transition ended at the faintest end of the hard state branch.  In the third (2007), however, it moved up and down the hard-state branch (never to the faintest end) where it remained during 2008/2009.  In future observations of GX 339-4 we will closely monitor the state transitions in the O/IR to see whether the paths traced in the correlation space vary or otherwise with each outburst.  In particular, we will see if the rise during the hard state from the faintest optical state to the brightest follows the same path as the decay phase at the faintest end of the hard-state branch.  It will also be interesting to see whether the location of the soft-branch merge remains the same in future hard-to-soft transitions.

The soft state O/IR SED is consistent with a thermal flux source that is most likely a heated accretion disk.  We were unable to discern whether this source is irradiated or not.  The spectral slope between $V$ and $I$ bands during the hard state is $\alpha = 1.3 \pm 0.1$.  For the soft state $\alpha = 1.6 - 1.8 \pm 0.1$, slightly higher than the hard-state optical slope, and consistent with those found by \cite{hyn05} and \cite{zur11} in other BHXBs ($0.5 \lesssim \alpha \lesssim 1.5$).  \cite{cor09} measured the spectral slope between the $H$- and $V$-band data to be $1 \le \alpha \le 2$ during the soft X-ray state and found it to be positively correlated with $V$-band flux. We also see the O/IR slope increasing with increasing brightness during the soft state, however, they are consistent within our errors.    

The hard-state SED shows evidence of both a non-thermal and thermal component.  A power-law spectrum extrapolated from the radio data in the hard-state SED during the 2002 outburst is consistent with the near-infrared data but not the optical.  The slopes derived for the radio/near-IR spectrum ($\alpha = 0.14-0.17 \pm 0.1$) are consistent with other studies \citep{cor02,hom05}.  On 18 April 2002 the errors on the radio data are sufficiently small to allow us to place tight constraints on the non-thermal power-law slope.  We do not see evidence of a non-thermal power-law break at frequencies lower than the $J$ band, suggesting that the break is located at frequencies greater than or equal to the $J$ band.  This is in agreement with \cite{cor09} but contrary to what was observed by \cite{gan11}. During 2010 when GX 339-4 was once again very bright during the hard state ($V$ = 14.74 mag) \cite{gan11} found the power-law break to be located in the mid-IR.  They fit a power-law to the radio-W4 data with a slope of +0.29 $\pm$ 0.02, significantly higher than slopes found in this paper and others.  Since we do not have simultaneous mid-IR data for the 2002 outburst, we cannot comment on what differences or similarities may exist in this spectral region.  However, if one considered just the 2010 radio data alone, the errorbar on the 9 GHz is sufficiently large that the radio spectrum could lie well below the mid-IR data and line up with the near-IR instead.  As Gandhi et al. also note, fitting single power-laws in the radio-to-near-IR spectral region may be an oversimplification.  Using sophisticated theoretical models, such as those shown in \cite{mar05}, \cite{hom05}, and \cite{mai09} may be necessary to explain the connection between the radio/mid-IR/near-IR data.  

\begin{figure}
\plottwo{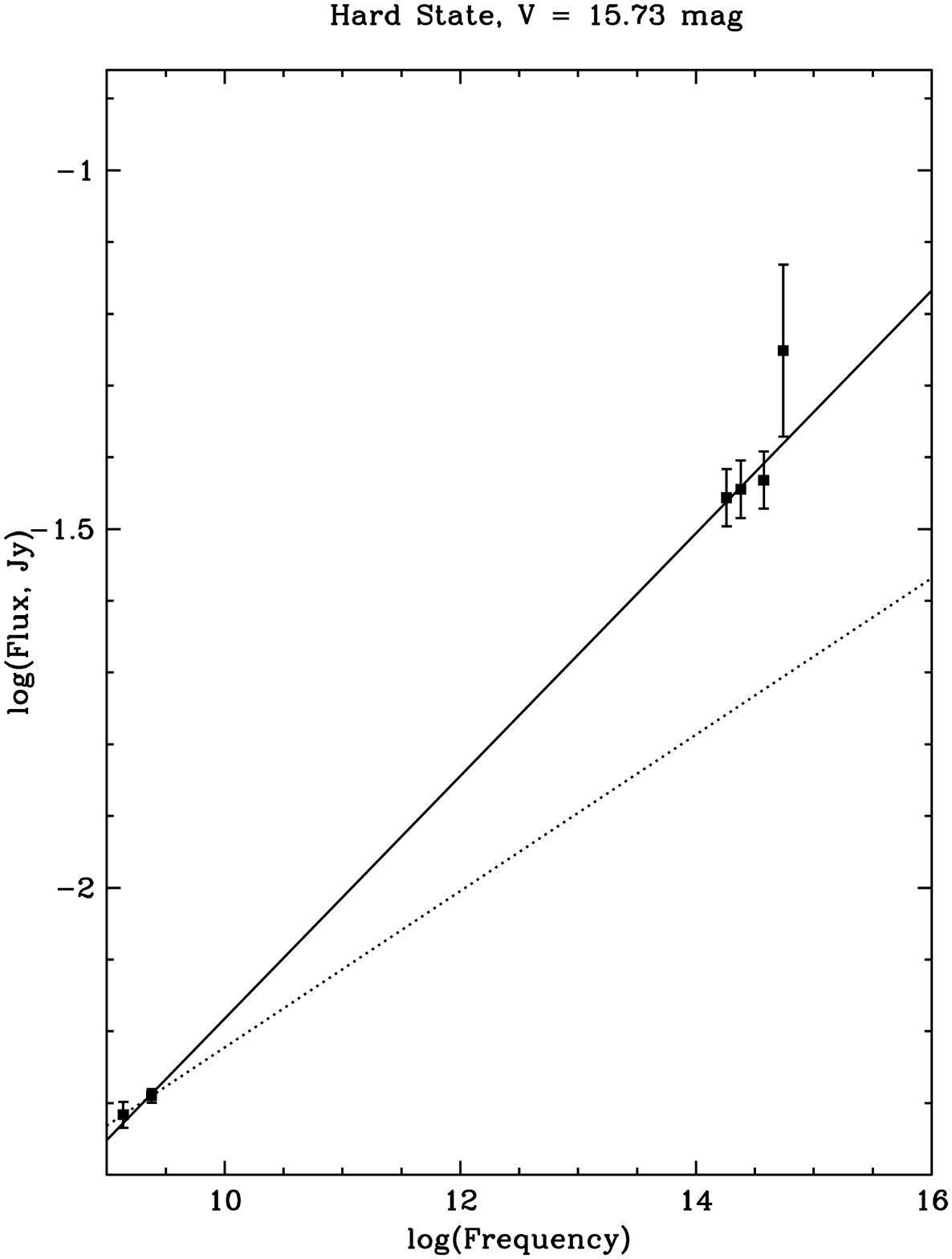}{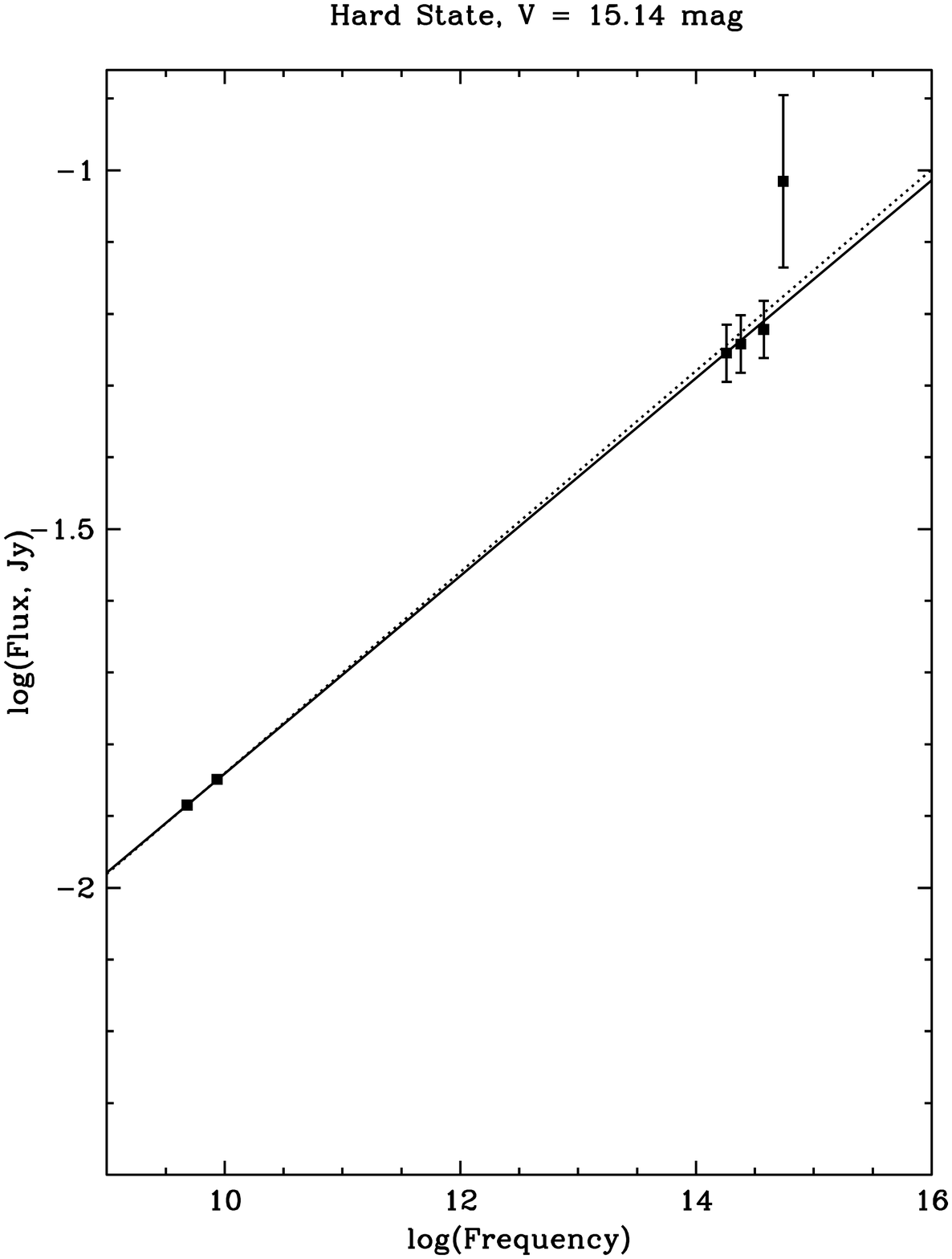}
\caption{Power-law model fits to the hard-state SED for $V$ = 15.73 ($left$) and 15.14 ($right$), when simultaneous O/IR and radio data were obtained on UT 03 and 18 April 2002, respectively.  The dotted line is a power-law fit to the radio data only ($left$: $\alpha$ = 0.11 $\pm$ 0.10; $right$ $\alpha$ = 0.14 $\pm$ 0.01).  The error on the radio data on 03 April 2002 is large enough such that the extrapolation may lie well above and below the OIR data.  On 18 April 2002, the radio spectrum lines up with the near-IR.  The solid line is a power-law fit to the radio, $J$ and $H$-bands ($left$: $\alpha$ = 0.18 $\pm$ 0.01; $right$ $\alpha$ = 0.17 $\pm$ 0.01). In this case, the radio and near-IR line up on the same spectral slope (within errors) on both dates and, therefore, originate from the same flux source, most probably jets.  The $V$ and $I$ bands are more consistent with a thermal flux source (as shown in Figure \ref{hardsed1}).  }
\label{hardsed2}
\end{figure}

\cite{cor09} used as yet unpublished radio and our SMARTS O/IR data to construct an SED when GX 339-4 was transitioning from the soft state to the hard state (UT 2005 Apr 21, MJD 53482).  The O/IR fluxes lie well $\it{below}$ the radio spectrum of slope $\alpha = 0.13 \pm$ 0.02.  Eleven days later (MJD 53493), the radio spectrum lines up with the near-IR with a slope $\alpha$ = 0.17 $\pm$ 0.01.  At this time, GX 339-4 reached its peak in the IR flux while the radio flux was declining.  Nine days later (MJD 53502) the radio flux decreased further while the O/IR flux decreased slightly and the near-IR was consistent with a radio spectrum of slope $\alpha$ = 0.21 $\pm$ 0.02.  Hence the power-law spectrum was significantly different during a soft-to-hard state transition than during the hard state during a rebrigtening phase.  Note this was at fainter O/IR mags than our SEDs in Figure \ref{hardsed2}.  

To model the hard-state O/IR SED would require a greater number of parameters than the data we have at hand can constrain.  As noted above, more sophisticated model-fits of the hard-state SED have been performed by \cite{mar05}, \cite{hom05}, and \cite{mai09}.  \cite{mar05} included data from the radio and X-ray regimes, and used quasi-simultaneous O/IR data obtained in 1981 by \cite{mot81} and \cite{ped81}. Their conclusions were that the O/IR could be explained by a combination of a multi-temperature \cite{sha73} disk and a single-temperature blackbody representing the irradiation of the outer accretion disk by the jet, with the IR dominated by flux from a jet.  \cite{hom05} applied models to simultaneous SMARTS O/IR ($V,I$, and $H$ bands), radio and X-ray data taken during the 2002 outburst.  Two cases were considered:  a pure-jet model (that is, no contribution to the optical by a disk), and one where the $V$ and $I$ band flux originate from a multicolor thermal disk and the $I$ to $H$ bands from a jet.  In both cases, there were difficulties in fitting the models to the steep $V$-to-$I$ band slope.  We note that in \cite{hom05} the $I$-band was not dereddened correctly.  In our paper we have corrected this and, as a result, we would see the $I$-band move up in the O/IR SED, producing a shallower slope between the $V$ and $I$ bands that would ease any difficulties Homan et al. had in fitting the models.  \cite{mai09} also used the same O/IR data as \cite{hom05} and concluded that their radio/O/IR/X-ray SED could be entirely explained by a jet-only model.  In light of the $I$-band dereddening correction, we strongly encourage a re-analysis of such model fitting.  In addition \cite{hom05} and \cite{mai09} used only three points \citep[$V$-, $I$-, and $H$-band][]{hom05}, whereas the 1981 dataset used by \cite{mar05} included 9 data points from $B$ to the near-IR.  Had data at shorter wavelengths been available, Homan et al. and Maitra et al. may have seen evidence for higher flux at shorter wavelengths (\cite{cad11} see a rise in flux from optical through to UV wavelengths) which may pose problems for a jet-only model fit.  

These results further highlight the importance of obtaining simultaneous radio data with the smallest errors and/or as many simultaneous bands as possible to place tight contraints on the radio spectral slope.  Sampling the optical/IR spectral region must now include the mid-IR and UV to sufficiently constrain theoretical models. Only then can we begin to have confidence in the location of the power-law break and how this varies over the duration of an outburst (i.e. with varying X-ray flux). 

Timing analysis during the soft ($V$-band) and hard states ($H$-band) show the variability of GX 339-4 is consistent with a simple power-law over timescales $\sim$4-230 days.  We see no evidence of characteristic timescales (such as power-law breaks or QPOs).  Other studies that have observed QPOs during the hard \citep{mot85,gan09,gan10} and soft \citep{mot85} state have been for timescales much less than what we've observed here ($\le$ 250 secs).  In future work we will obtain data with higher temporal resolution in an attempt to bridge the timescale-gap between 250 secs and 4 days during both hard and soft X-ray states, and construct PSDs, cross-, and auto-correlation functions to fully explore variability timescales in GX 339-4 to see any evidence of characteristic timescale(s) related to accretion processes.  

We will continue to monitor GX 339-4 in the O/IR with SMARTS for the foreseable future, including data at $B$-, $V$-, $I$-, $J$- $H$-, and $K$-bands, making it immediately public via the web\footnote{http://www.astro.yale.edu/buxton/GX339/} for use by the general astronomical community.

\acknowledgments

We are grateful to our SMARTS ANDICAM observers, Juan Espinoza, Alberto Miranda, and David Gonzalez, and to Suzanne Tourtellotte for reducing the optical data.  
  
MMB, CDB, HC, and RC are supported by NSF/AST grants 0407063 and 070707 to CDB.

J.A.T. acknowledges partial support from NASA under grant number NNX09AG46G.

EK and TD acknowledge TUBITAK grant 106T570,  Turkish Academy of Sciences and the EU FP7, 
ITN 215212 "Black Hole Universe".

This research has made use of the NASA/ IPAC Infrared Science Archive, which is operated by the 
Jet Propulsion Laboratory, California Institute of Technology, under contract with the National 
Aeronautics and Space Administration.

{\it Facilities:} \facility{CTIO SMARTS (1.3m)}.

\clearpage

\clearpage
\begin{deluxetable}{cccccccc}
\tablecaption{Optical comparison star magnitudes. Errors are 1-$\sigma$. \label{compstarsop}}
\tablehead{
\colhead{Number} & \colhead{RA} & \colhead{Dec} & \colhead{$V$} & \colhead{$I$} & \colhead{($V-I$)}\\
 & \colhead{(J2000.0)} & \colhead{(J2000.0)} & \colhead{(mag)} & \colhead{(mag)} & \colhead{(mag)} }
\startdata
1 & 17:02:52.0 & -48:47:09.5 & 17.54 $\pm$ 0.06 & 15.26 $\pm$ 0.07 & 2.28 \\
2 & 17:02:49.9 & -48:47:17.1 & 17.43 $\pm$ 0.05 & 15.49 $\pm$ 0.07 & 1.94 \\
3 & 17:02:51.8 & -48:47:21.8 & 16.99 $\pm$ 0.05 & 15.62 $\pm$ 0.06 & 1.37 \\
\enddata
\tablecomments{Star numbers are in accordance with the optical finding chart (Figure \ref{opfc}).}
\end{deluxetable}

%

\clearpage
\begin{deluxetable}{ccccc}
\tablecaption{Near-infrared comparison star magnitudes. Errors are 1-$\sigma$. \label{compstarsir}}
\tablehead{
\colhead{Number} & \colhead{RA} & \colhead{Dec} & \colhead{$J$} & \colhead{$H$} \\
 & \colhead{(J2000.0)} & \colhead{(J2000.0)} & \colhead{(mag)} & \colhead{(mag)}}
\startdata
1 & 17:02:52.0 & -48:47:09.5 & 13.35 $\pm$ 0.09 & 12.49 $\pm$ 0.09 \\
2 & 17:02:49.9 & -48:47:17.1 & 13.88 $\pm$ 0.09 & 13.02 $\pm$ 0.10 \\
\enddata
\tablecomments{Star numbers are in accordance with the IR finding chart (Figure \ref{irfc}).  Our magnitudes are consistent with the 2MASS catalog within our photometric errors.}
\end{deluxetable}


\clearpage
\begin{deluxetable}{cccccccccccccccccc}
\tabletypesize{\tiny}
\rotate
\tablecaption{Calibrated $V$-, $I$-, $J$-, and $H$-band magnitudes, fluxes and their errors of GX 339-4 from 2002 - 2010, inclusive (abbreviated).  The JD is that at the start of the observing sequence.  The full version can be accessed online.\label{alldata}}
\tablewidth{0pt}
\tablehead{
\colhead{year} & \colhead{JD} & \colhead{$V$ mag} & \colhead{$V$ err} & \colhead{$V_0$ flux} & \colhead{$V_0$ err} & \colhead{$I$ mag} & \colhead{$I$ err} & \colhead{$I_0$ flux} & \colhead{$I_0$ err} & \colhead{$J$ mag} & \colhead{$J$ err} & \colhead{$J_0$ flux} & \colhead{$J_0$ err} & \colhead{$H$ mag} & \colhead{$H$ err} & \colhead{$H_0$ flux} & \colhead{$H_0$ err} \\
 & (-2450000.0) & \colhead{(mag)} & \colhead{(mag)} & \colhead{(Jy)} & \colhead{(Jy)} & \colhead{(mag)} & \colhead{(mag)} & \colhead{(Jy)} & \colhead{(Jy)} & \colhead{(mag)} & \colhead{(mag)} & \colhead{(Jy)} & \colhead{(Jy)} & \colhead{(mag)} & \colhead{(mag)} & \colhead{(Jy)} & \colhead{(Jy)}
}
\startdata
2002 &   2297.85400 & - & -   & -   & - & 17.730 & 0.030  &    0.00103  &  0.00010   &  - & -   & - & -  & 16.080  & 0.070   &  0.00069   & 0.00007 \\
     &   2298.87070 & - & -   & -   & - & -      &   -    &      -      &     -      &  - & -   & - &  - & 16.640  & 0.110   &  0.00041   & 0.00004 \\
     &   2314.84330 & - & -   & -   & - & 18.280 &  0.040 &    0.00062  &  0.00006   &  - & -   & - &  - & 16.360  & 0.090   &  0.00053   & 0.00005 \\
     &   2314.84830 & - & -   & -   & - & -      &   -    &      -      &      -     &  - & -   & - &  - & 16.340  & 0.090   &  0.00054   & 0.00005 \\
     &   2316.83520 & - & -   & -   & - & -      &   -    &      -      &      -     &  - & -   & - &  - & 16.480  & 0.100   &  0.00048   & 0.00005 \\
     &   2317.84150 & - & -   & -   & - & 17.840 &  0.030 &    0.00093  &  0.00009   &  - & -   & - &  - & 16.470  & 0.100   &  0.00048   & 0.00005 \\
     &   2318.81040 & - & -   & -   & - & 17.630 &  0.030 &    0.00112  &  0.00011   &  - & -   & - &  - & 15.830  & 0.060   &  0.00087   & 0.00008 \\
     &   2319.85140 & - & -   & -   & - & 17.570 &  0.030 &    0.00119  &  0.00011   &  - & -   & - &  - & 16.210  & 0.080   &  0.00061   & 0.00006 \\
     &   2320.83940 & - & -   & -   & - & 17.600 &  0.030 &    0.00116  &  0.00011   &  - & -   & - &  - & 16.090  & 0.070   &  0.00068   & 0.00007 \\
     &   2322.86770 & - & -   & -   & - & 17.440 &  0.020 &    0.00134  &  0.00013   &  - & -   & - &  - & 15.710  & 0.060   &  0.00097   & 0.00009 \\

\enddata
\end{deluxetable}

\clearpage
\begin{deluxetable}{ccccc}
\tablecaption{Slopes of best-fit lines to 2003 and 2005 light curves (see Figure \ref{rebrights}) immediately before the rebrightening events.  Errors stated are 1-$\sigma$. \label{rebrights_slopes}}
\tablehead{
\colhead{Year} & \colhead{$V$} & \colhead{$I$} & \colhead{$J$} & \colhead{$H$} \\
 & \colhead{(mag/day)} & \colhead{(mag/day)} & \colhead{(mag/day)} & \colhead{(mag/day)}}
\startdata
2003 & 0.010 $\pm$ 0.002 & 0.008 $\pm$ 0.002 & 0.007 $\pm$ 0.003 & 0.003 $\pm$ 0.002 \\
2005 & 0.012 $\pm$ 0.001 & 0.010 $\pm$ 0.001 & 0.009 $\pm$ 0.001 & 0.007 $\pm$ 0.001 \\
\enddata
\end{deluxetable}

\clearpage
\begin{deluxetable}{ccccc}
\tablecaption{Power-law fit parameters to O/IR correlations shown in Figure \ref{oircorr}.  Soft-state branch requires only a single power-law fit ($F_y = C.F_x^{\alpha_1}$).  The hard-state branch requires a broken power-law ($F_y = C.F_x^{\alpha_1}$ for $x \le x_{break}$, and $F_y = C.F_x^{\alpha_1}.F_{x(break)}^{\alpha_1 - \alpha_2}$ for $x > x_{break}$). Errors are 1-$\sigma$. \label{corfit}}
\tablehead{
\colhead{Data/Branch} & \colhead{$\alpha_1$} & \colhead{$\alpha_2$} & \colhead{$x_{break}$} & \colhead{C} \\
                      &                      &                      & (mag)                 & (mag)
}
\startdata
$V/I$ hard & 1.039 $\pm$ 0.001 & 0.890  $\pm$ 0.005 & 16.07 $\pm$ 0.03 & -2.06 $\pm$ 0.02 \\
$V/J$ hard & 1.242 $\pm$ 0.002 & 0.858  $\pm$ 0.006 & 16.20 $\pm$ 0.01 & -7.04 $\pm$ 0.04 \\
$V/H$ hard & 1.362 $\pm$ 0.002 & 0.853  $\pm$ 0.005 & 16.30 $\pm$ 0.01 & -9.89 $\pm$ 0.04 \\
$V/I$ soft & 0.893 $\pm$ 0.002 & n/a & n/a & 0.71 $\pm$ 0.03 \\
$V/J$ soft & 0.796 $\pm$ 0.004 & n/a & n/a & 1.41 $\pm$ 0.06 \\
$V/H$ soft & 0.722 $\pm$ 0.003 & n/a & n/a & 2.25 $\pm$ 0.06 \\
\enddata
\end{deluxetable}

\clearpage
\begin{deluxetable}{cccccc}
\tablecaption{A summary of the best-fit power-law parameters to the power spectral densities of GX 339-4 during the hard and soft states. Errors are 1-$\sigma$. \label{psdfits}}
\tablehead{
\colhead{Filter/X-ray state} & \colhead{Start JD} & \colhead{End JD} & \colhead{Slope ($\alpha$)}\\
                             & (-2450000.0)       & (-2450000.0)     &                    
}
\startdata
$V$ soft state & 3224.70050 & 3478.79434 & $-2.0^{+0.3}_{-0.5}$ \\
$V$ hard state & 4498.88162 & 4746.54995 & $-1.8^{+0.3}_{-0.5}$  \\
$H$ soft state & 3224.70050 & 3478.79434 & $-1.7^{+0.2}_{-0.6}$  \\
$H$ hard state & 4498.88162 & 4746.54995 & $-1.8^{+0.3}_{-0.7}$  \\
\enddata
\end{deluxetable}

\end{document}